\documentclass[12pt]{article}
\usepackage{geometry}

\usepackage{amsmath, amssymb, graphicx,fullpage,color,mathtools,amsthm,xcolor}
\usepackage{caption, subcaption}
\usepackage{algorithm, algorithmic}
\usepackage{cite}
\usepackage{subcaption}
\usepackage{array}
\usepackage{url}
\usepackage{tikz}
\usepackage{tabularx}
\usepackage{booktabs}
\usetikzlibrary{positioning,fit}
\usepackage[normalem]{ulem}

\usetikzlibrary{positioning,fit}

\usepackage{pythonhighlight}

\usepackage[table]{xcolor} 
\definecolor{Gray}{gray}{0.9} 

\usepackage{pifont}

\usepackage{microtype}
\usepackage{hyperref,color}

\definecolor{webgreen}{rgb}{0,.35,0}
\definecolor{webbrown}{rgb}{.6,0,0}
\definecolor{RoyalBlue}{rgb}{0,0,0.9}
\definecolor{purp}{rgb}{0.6,0.05,0.8}
\definecolor{ora}{rgb}{0.7,0.35,0.02}

\usepackage{pythonhighlight}

\usepackage[table]{xcolor} 
\definecolor{Gray}{gray}{0.9} 

\usepackage{multirow}
\newcolumntype{C}[1]{>{\centering\let\newline\\\arraybackslash\hspace{0pt}}m{#1}}

\hypersetup{
   colorlinks=true, linktocpage=true, 
   urlcolor=webbrown, linkcolor=RoyalBlue, citecolor=webgreen,
   pdfauthor={Qinghai Jiang, Gary P. T. Choi},
   pdfsubject={PyKirigami: An interactive Python simulator for kirigami structures}
}

\begin{document}

\author{Qinghai Jiang$^{1}$, Gary P. T. Choi$^{1,\ast}$\\
\\
\footnotesize{$^{1}$Department of Mathematics, The Chinese University of Hong Kong}\\
\footnotesize{$^\ast$To whom correspondence should be addressed; E-mail: ptchoi@cuhk.edu.hk}
}
\title{PyKirigami: An interactive Python simulator for kirigami structures}
\date{}
\maketitle

\begin{abstract}
In recent years, the concept of kirigami has been used in creating deployable structures for various scientific and technological applications. While high-fidelity Finite Element Analysis (FEA) is the standard for analyzing stress distributions and material deformation, it is computationally intensive and often ill-suited for the rapid exploration of vast kinematic configuration spaces. In this work, we develop PyKirigami, a lightweight, open-source Python framework for the efficient deployment simulation of kirigami structures. Unlike continuum mechanics solvers, PyKirigami models tessellations as articulated rigid-body networks, allowing for the real-time simulation of global deployment trajectories and volumetric transformations. The tool incorporates collision detection and interactive actuation, enabling users to validate folding paths and identify geometric locking states in both 2D and 3D topologies. This framework serves as a fast kinematic prototyping tool for kirigami structures, allowing researchers to verify deployment mechanics and self-contacts prior to performing detailed mechanical analysis or physical fabrication.
\end{abstract}

\section{Introduction}
Kirigami, the traditional paper-cutting art, has recently become a subject of great interest in science and engineering~\cite{barchiesi2019mechanical,zhai2021mechanical,li2023auxetic,choi2024computational,jin2024engineering,dang2025kirigami}. In particular, a wide range of deployable structures with different shape-morphing effects have been developed by introducing cuts on a sheet of materials and connecting different components using suitable joints. Because of their great flexibility and shape-shifting property, kirigami-based structures have been widely applied in soft electronics~\cite{jiang2022flexible}, energy storage~\cite{song2015kirigami}, and robotics~\cite{yang2021grasping}.

Most prior works on kirigami design have focused on planar deployable structures and their in-plane deformations based on simple cutting patterns, such as rotating triangles~\cite{grima2006auxetic}, quadrilaterals~\cite{grima2000auxetic,attard2008auxetic}, and other periodic tilings~\cite{rafsanjani2016bistable,stavric2019geometrical,liu2021wallpaper} and aperiodic tilings~\cite{liu2022quasicrystal}, as well as their applications to the design of graphene kirigami with high stretchability and ductility~\cite{shi2023deformation} and crystallographically programmed metamaterials~\cite{he2024crystallographically}. In recent years, some works have also explored the design of more general kirigami patterns for achieving different physical and geometrical effects~\cite{chen2016topological,choi2019programming,chen2020deterministic,choi2021compact,dang2021theorem,zheng2022continuum,hong2022boundary,wang2023physics,dudte2023additive,qiao2025inverse}. There is also an increasing interest in designing kirigami for achieving not just two-dimensional (2D) but three-dimensional (3D) shape transformations with different desired 2D-to-3D or 3D-to-3D shape-morphing effects~\cite{sussman2015algorithmic,konakovic2016beyond,konakovic2018rapid,wang2020keeping,chen2021bistable,dang2022theorem,li20213d,dang2025shape}, together with their morphological and mechanical performance~\cite{he2025programming}.

While the design of kirigami structures has been extensively studied, computational methods for the simulation and analysis of their shape-morphing process are much less explored. For origami structures, many existing computational tools have been established for simulating their folding process. For instance, the Rigid Origami Simulator~\cite{tachi2009simulation} simulates the kinematics of rigid origami from given crease patterns. The Freeform Origami~\cite{tachi2010freeform} simulates origami folding and provides interactive functionalities for altering the model crease patterns. MERLIN~\cite{liu2016merlin} and MERLIN2~\cite{liu2018highly} are MATLAB-based software tools for simulating the folding of non-rigid origami. Origami Simulator~\cite{ghassaei2018fast} is a GPU-accelerated and web-based origami simulator with interactive folding control and real-time rendering. On the contrary, for kirigami structures, most prior kirigami design works directly used finite element analysis (FEA) software such as ABAQUS (e.g.,~\cite{jin2020kirigami,an2020programmable,qiao2025inverse}) or COMSOL (e.g.,~\cite{zhu2018kirigami,alderete2021programmable,ying2025inverse}) to simulate the deployment process of kirigami structures under certain boundary loading conditions. Some other works considered other mechanical models involving linear springs for simulating 2D-to-2D and 2D-to-3D deployments in MATLAB~\cite{choi2019programming,choi2021compact}. More recently, Liu et al.~\cite{liu2022quasicrystal} developed a 2D-to-2D kirigami deployment simulator in Python based on the 2D rigid body physics library Pymunk, with a focus on rigidly deployable planar kirigami patterns. However, in many practical applications, it is necessary to simulate more general kirigami deployment processes in both 2D and 3D. Also, one may need to consider not just the kirigami structures but also their interactions with the environment and external objects throughout the deployment. Moreover, it is highly desirable to have a lightweight, efficient, and freely available deployment simulator with interactive control functionalities that allow for real-time manipulation and analysis of kirigami structures.

\begin{figure}[t!]
    \centering
    \includegraphics[width=\linewidth]{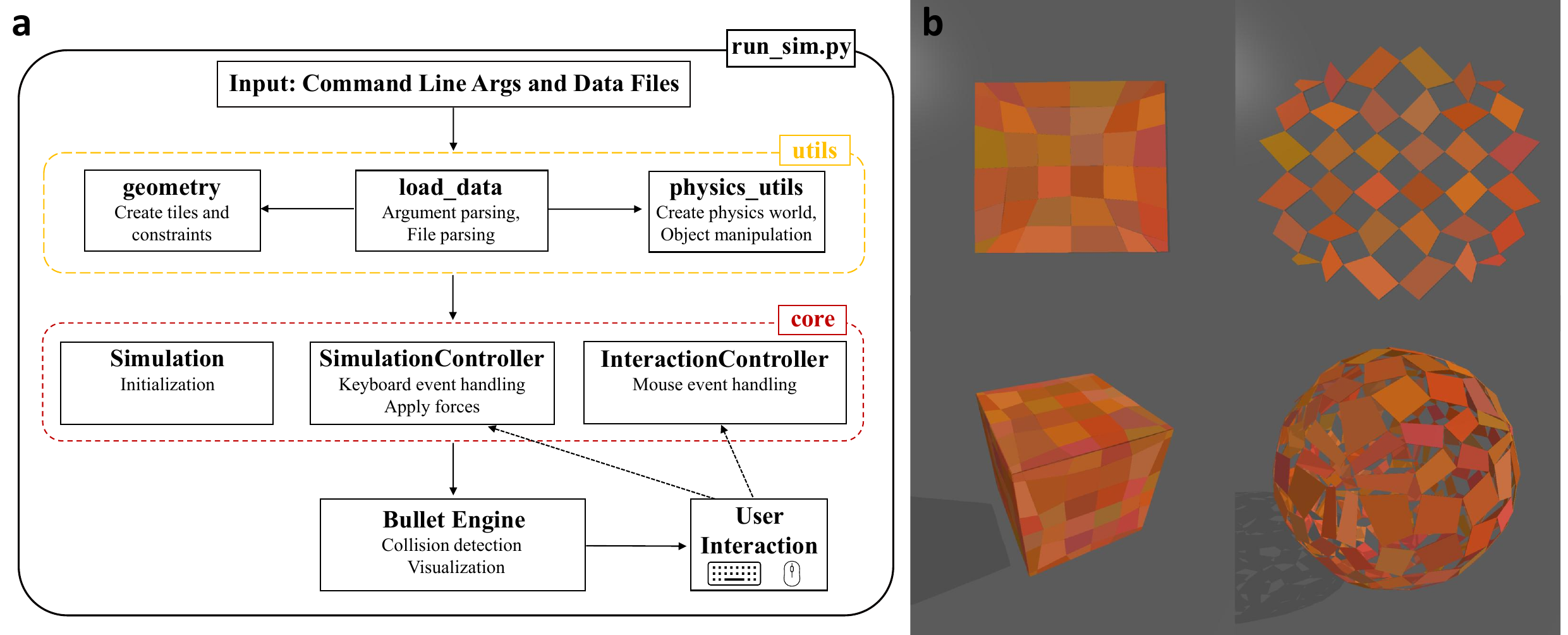}
    \caption{\textbf{An overview of PyKirigami.} \textbf{a}, The computational framework of PyKirigami. \textbf{b}, Using PyKirigami, one can easily simulate the deployment process of different kirigami structures in 2D (top) and 3D (bottom).}
    \label{fig:overview}
\end{figure}

In this work, we develop PyKirigami, an open-source Python-based simulator, for simulating the 2D and 3D deployment of general kirigami structures (Fig.~\ref{fig:overview}). Specifically, our simulator utilizes the PyBullet physics engine~\cite{coumans2021} and is capable of simulating 2D-to-2D, 2D-to-3D, and 3D-to-3D kirigami deployments. It also supports the configurations of environmental effects and stabilization mechanisms for modelling different scenarios. It further allows dynamic control of the boundary conditions throughout the deployment process, thereby providing great flexibility in achieving different desired deployment effects. Using PyKirigami, one can easily simulate the deployment and analyze the properties of a wide range of kirigami structures.

\section{Methods} 

\subsection{Overview}
The goal of our PyKirigami simulator is to simulate the deployment process of kirigami structures, typically from a compact initial state to a deployed configuration in either 2D or 3D. More specifically, the functionalities of PyKirigami (Fig.~\ref{fig:overview}\textbf{a}) can be categorized as follows:
\begin{itemize}
    \item \textbf{Core Simulation Logic}: This component is responsible for the main simulation loop, handling physics calculations, and managing the state of the simulated objects. It includes classes for managing the simulation state, handling user interactions (like pausing and resetting), and applying forces to the objects.
    \item \textbf{Utility Functions}: This group contains modules for parsing command-line arguments, loading input data (such as vertex coordinates and constraints), and performing geometric operations like creating 3D models from 2D definitions.
    \item \textbf{Main Executable}: The main executable script initializes the simulation environment, orchestrates the simulation loop, and integrates the various modules.
\end{itemize}

The core concepts in PyKirigami are presented in further detail in the following sections. See also Appendix~\ref{appendix:overview}--\ref{appendix:implementation} for more detailed explanations.

\subsection{Geometric Representation: From 2D Polygons to 3D Tiles}
In our work, we consider polygonal tiles formed by straight cuts, noting that curved cuts can be discretized and approximated using multiple straight cuts in practical applications. Also, in prior works on kirigami design and simulation~\cite{choi2019programming,liu2022quasicrystal}, the kirigami tiles were commonly considered as two-dimensional objects with zero thickness for simplicity. By contrast, here we consider the kirigami tiles as 3D rigid bodies (``bricks'') in both the in-plane and 3D deployment simulations, thereby enabling a more realistic simulation process. 

\begin{figure}[t]
    \centering
    \includegraphics[width=\linewidth]{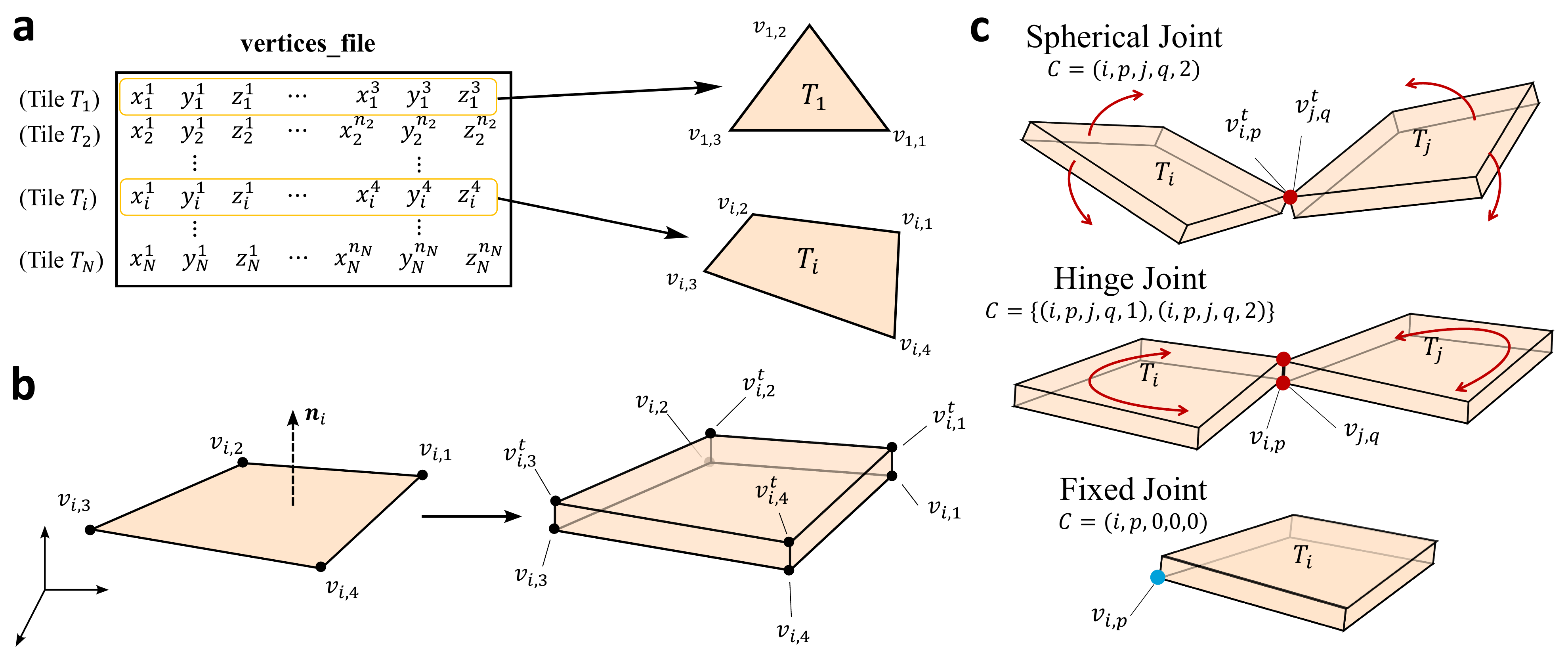}
    \caption{\textbf{Tile construction and connections in PyKirigami.} \textbf{a}, An illustration of the vertex coordinates encoded in the input file via \texttt{--vertices\_file}. Here, note that different tiles may contain different number of vertices. \textbf{b}, Geometric construction of the 3D kirigami tiles. (Left) The original polygonal tile $\mathcal{T}_i$, with the dotted arrow showing its normal vector. (Right) The 3D tile geometry constructed by extruding $\mathcal{T}_i$ along its normal vector. Here, the thickness of the tile can be prescribed by the user. \textbf{c}, Three different types of inter-tile connections in PyKirigami. (Top) Spherical joint applied to tiles $\mathcal{T}_{i}$ and $\mathcal{T}_{j}$ such that the rotational degree of freedom equals $3$. (Middle) Hinge joint such that the rotational degree of freedom equals $1$. (Bottom) Fixed joint applied to $\mathcal{T}_{i}$ and world frame such that the object is totally fixed. The arrow represents the rotational degree of freedom.}
\label{fig:illustration_tile_connection_combined}
\end{figure}

More mathematically, let $\{\mathcal{T}_i\}_{i=1}^N$ be the collection of all kirigami tiles in a kirigami structure, where each tile $\mathcal{T}_i$ is initially defined by a sequence of vertices $\{v_{i,1}, v_{i,2}, \dots, v_{i,n_i}\}$, where $v_{i,j} \in \mathbb{R}^3$ and $n_i$ is the number of vertices for tile $\{\mathcal{T}_i\}$ (see also Fig.~\ref{fig:illustration_tile_connection_combined}\textbf{a}). To create a more physically realistic model, each planar tile is extruded into a 3D rigid body, or ``brick," with a user-defined thickness, $t$. This process is detailed in Algorithm~\ref{alg:tile_generation} and illustrated in Fig.~\ref{fig:illustration_tile_connection_combined}\textbf{b}. See also Appendix~\ref{appendix:implementation} for more details of the geometric representation.

\begin{algorithm}[H]
\caption{3D Tile Generation Pipeline}
\label{alg:tile_generation}
\begin{algorithmic}[1]
\STATE \textbf{Input:} Vertex coordinates file, brick thickness parameter $t$
\STATE \textbf{Output:} PyBullet body IDs, local vertex coordinates for each tile

\STATE Load vertex data from input file
\FOR{each tile $\mathcal{T}_i$ with vertices $\{v_{i,1}, v_{i,2}, \ldots, v_{i,n_i}\}$}
    \STATE Compute geometric center: $\mathbf{c}_i = \frac{1}{n_i}\sum_{j=1}^{n_i} v_{i,j}$.
    \STATE Calculate normal vector: $\mathbf{n}_i = \frac{(v_{i,2} - v_{i,1}) \times (v_{i,3} - v_{i,1})}{\|(v_{i,2} - v_{i,1}) \times (v_{i,3} - v_{i,1})\|}$.
    \STATE Generate top vertices: $v_{i, j}^{t} = v_{i,j} + t \cdot \mathbf{n}_i$ for $j = 1, \ldots, n_i$.
    \STATE Store local coordinates: $v_{i, j} - \mathbf{c}_{i}$ and $v_{i, j}^{t}-\mathbf{c}_{i}$.
    \STATE Create visual mesh and collision shape from $2n_i$ vertices.
    \STATE Instantiate rigid body in physics engine, return body ID.
\ENDFOR
\end{algorithmic}
\end{algorithm}

We remark that by modelling the polygonal tiles as 3D rigid bodies, PyKirigami focuses on the rigid deployment of kirigami structures and excludes other complex effects such as panel bending, stretching, crease compliance, material elasticity, plasticity, and fracture. Certain non-rigid deployment scenarios may be further handled by some specific strategies and simulation setups, which are discussed in Appendix~\ref{appendix:complex}.

\subsection{Kinematic Connections: Modeling Joints with Constraints} \label{sect:connections}
The connections between tiles in kirigami structures are crucial for their kinematic behavior. Building upon existing work, we adopt and extend the constraint representation framework introduced in~\cite{liu2022quasicrystal}.

\paragraph{Extension from 2D to 3D Constraint Representation}
In the prior 2D work~\cite{liu2022quasicrystal}, each constraint is represented as a 4-tuple $\mathcal{C}_k = (i, p, j, q)$, which enforces that vertex $p$ of tile $i$ coincides with vertex $q$ of tile $j$. This constraint effectively creates a spherical joint in 2D space such that
\begin{equation}
    v_{i,p} = v_{j,q}.
\end{equation}
Extending this to 3D naturally requires an additional $z$-coordinate constraint. However, since our tiles are 3D rigid bodies with distinct top and bottom faces, we must specify which face contains the connection point. We therefore extend the constraint representation to a 5-tuple $\mathcal{C}_k = (i, p, j, q, t)$, where $t$ is a face specifier: $t = 1$ for bottom face connections and $t = 2$ for top face connections. For backward compatibility, if $t$ is omitted (4-tuple format), we default to bottom face connection.

\paragraph{Joint Types and Their Implementation}
By strategically combining different joints in PyBullet, we can realize different types of kinematic connections with varying degrees of freedom. Each connection type corresponds to a specific pattern of vertex coincidence constraints (see Fig.~\ref{fig:illustration_tile_connection_combined}\textbf{c}):

\begin{itemize}
    \item \textbf{Spherical Joint}: The fundamental connection type, implemented using a single point-to-point constraint that enforces coincidence of one vertex from each tile (Fig.~\ref{fig:illustration_tile_connection_combined}\textbf{c}, top). This creates a ball-and-socket joint allowing three rotational degrees of freedom around the pivot point.

    \item \textbf{Hinge Joint}: Another connection type constructed using two spherical joints to constrain rotation to a single axis (Fig.~\ref{fig:illustration_tile_connection_combined}\textbf{c}, middle). 

    \item \textbf{Fixed Joint}: A connection type implemented using PyBullet's native \texttt{JOINT\_FIXED} constraint type (Fig.~\ref{fig:illustration_tile_connection_combined}\textbf{c}, bottom). This rigid connection completely eliminates all relative degrees of freedom between objects, preventing any relative motion. Fixed joints are particularly useful for creating structural stability and for temporarily immobilizing specific tiles during interactive simulations.

\end{itemize}

We remark that besides utilizing the user-provided connection information, PyKirigami also provides an automatic connection-type detection functionality for users who are unsure about the connection types to be used. See~\ref{subsec:constraint} for more details.

\subsection{Force Models for Actuation and Dynamics} \label{sect:dynamics}
After constructing the kirigami tiles and their connections, we implement a physics-based deployment simulation. The dynamic behavior of the kirigami structure is governed by various configurable force models, including simple expansion forces for basic radial deployment and more advanced target-based actuation.

\paragraph{Center-of-Mass Expansion Forces}
For basic radial expansion of kirigami structures, PyKirigami provides an automatic expansion mode that applies forces to each tile's center of mass, directing them outward from the structure's global center. This method is particularly useful when the target positions are not specified, serving as a fallback or simplified actuation approach. 

\paragraph{Target-Based Deployment Forces}
While simple radial expansion can be achieved using center-of-mass forces described above, more complex deployments require a more sophisticated approach. Our simulator implements a vertex-based target model that provides precise control over the deployment process. In this model, each vertex of each tile is assigned a target position, and forces are applied to guide the vertices toward these targets. The deployment process is implemented through Algorithm~\ref{alg:target_forces}, which calculates and applies appropriate forces and torques to achieve the desired configuration:

\begin{algorithm}[H]
\caption{Vertex-Based Target Deployment Framework}
\label{alg:target_forces}
\begin{algorithmic}[1]
\STATE \textbf{Input:}
    Body IDs $\{b_i\}$;
    local vertex coordinates $\{\mathbf{u}_{i,j}\}_{j=1}^{m_i}$ per tile;
    target vertex positions $\{\tilde{v}_{i,j}\}_{j=1}^{m_i}$;
    spring stiffnesses $k_f$, $k_\tau$;
    damping coefficients $\mu_v$, $\mu_\omega$

\FOR{each body $b_i$ representing tile $\mathcal{T}_i$}
    \STATE Query centre of mass $\mathbf{c}_i$, orientation $R_i$,
           linear velocity $\mathbf{v}_i$, and angular velocity
           $\boldsymbol{\omega}_i$ from the physics engine.
    \STATE Transform local vertices to world frame:
           $v_{i,j} = R_i\,\mathbf{u}_{i,j} + \mathbf{c}_i$.
    \STATE Initialise $\mathbf{F}_i \leftarrow \mathbf{0}$, $\boldsymbol{\tau}_i \leftarrow \mathbf{0}$.
    \FOR{$j = 1$ \TO $m_i$}
        \STATE Compute displacement:
               $\boldsymbol{\delta}_{i,j} = \tilde{v}_{i,j} - v_{i,j}$.
        \STATE Compute moment arm from CoM:
               $\mathbf{r}_{i,j} = v_{i,j} - \mathbf{c}_i$.
        \STATE $\mathbf{F}_i \mathrel{+}= k_f\,\boldsymbol{\delta}_{i,j}$
        \STATE $\boldsymbol{\tau}_i \mathrel{+}=
               k_\tau\,(\mathbf{r}_{i,j} \times \boldsymbol{\delta}_{i,j})$
    \ENDFOR
    \STATE Apply damped force and torque to body $b_i$:
    \[
        \hat{\mathbf{F}}_i = \mathbf{F}_i - \mu_v\,\mathbf{v}_i,
        \qquad
        \hat{\boldsymbol{\tau}}_i = \boldsymbol{\tau}_i - \mu_\omega\,\boldsymbol{\omega}_i.
    \]
\ENDFOR
\end{algorithmic}
\end{algorithm}

This vertex-based approach offers several advantages over simpler force models, including precise shape control, natural rotational behavior, and the ability to follow complex deployment paths.

\paragraph{Adaptive Stiffness}
We remark that an issue with the above-mentioned linear-spring actuation is that the force $k_f(\tilde{v}_{i,j} - v_{i,j})$ decays to zero as each vertex approaches its target, making it difficult to overcome residual friction or geometric locking. To address this, PyKirigami employs an adaptive stiffness scheme. During the automatic deployment, the controller monitors the maximum vertex error across all active tiles. If the error fails to improve for 60 consecutive simulation steps, the spring constant $k_f$ is doubled (up to a maximum of $\times 16$).

\paragraph{Environmental and Stabilization Forces} 
Beyond the actuation forces, PyKirigami models key environmental effects and stabilization mechanisms, which are fully configurable by the user to simulate diverse physical scenarios:
\begin{itemize}
\item \textbf{Gravitational Loading:} The magnitude and direction of the gravity vector can be specified to model deployment under various load conditions.
\item \textbf{Numerical Damping:} To enhance stability and simulate energy dissipation, the model includes two configurable damping mechanisms. First, tunable linear and angular damping coefficients are applied to each rigid body to dissipate system-wide kinetic energy and ensure stable convergence to quasi-static equilibrium. Second, the actuation forces incorporate a damping factor (as seen in Algorithm~\ref{alg:target_forces}) to promote smooth and stable deployment dynamics.
\item \textbf{Contact Modeling:} The simulator supports interactions with external surfaces, such as a configurable ground plane, and models the resulting collision and friction forces. 
\end{itemize}

\paragraph{Self-penetration Avoidance} PyKirigami tries to avoid self-penetration during the deployment simulation through several complementary mechanisms that operate at different levels of the simulation pipeline. In particular, besides the built-in rigid body collision detection offered in PyBullet, PyKirigami also includes functionalities for avoiding self-penetration potentially caused by the inter-tile connection type and incorrect global target assignment. It further provides functions for decomposing complex deployments into a sequence of simpler sub-deployments to avoid self-penetration due to complex target shape changes. A more detailed description is provided in Appendix~\ref{appendix:self_penetration}.

\section{Results} 
\subsection{2D and 3D Deployment Simulations}
To demonstrate the versatility and effectiveness of PyKirigami, we present a curated set of examples grouped by deployment dimensionality. These cases showcase how the simulator's automated deployment mechanisms, environmental controls, and interactive features are applied to solve distinct challenges in 2D and 3D kirigami systems. For detailed simulation parameters and per-example configurations, see Appendix~\ref{appendix:examples}. For the software environment and detailed commands for reproducing all examples, see Appendix~\ref{appendix:reproducibility}. See also Supplementary Video S1--S4 for the videos of the deployment process of various 2D and 3D kirigami examples.

\subsubsection{2D Deployments and Interactive Morphing}
Planar kirigami systems often involve in-plane expansion or complex shape-morphing between multiple stable states. We demonstrate PyKirigami's capabilities for these scenarios with two distinct examples.

\begin{figure}[t!]
    \centering
    \includegraphics[width=0.8\linewidth]{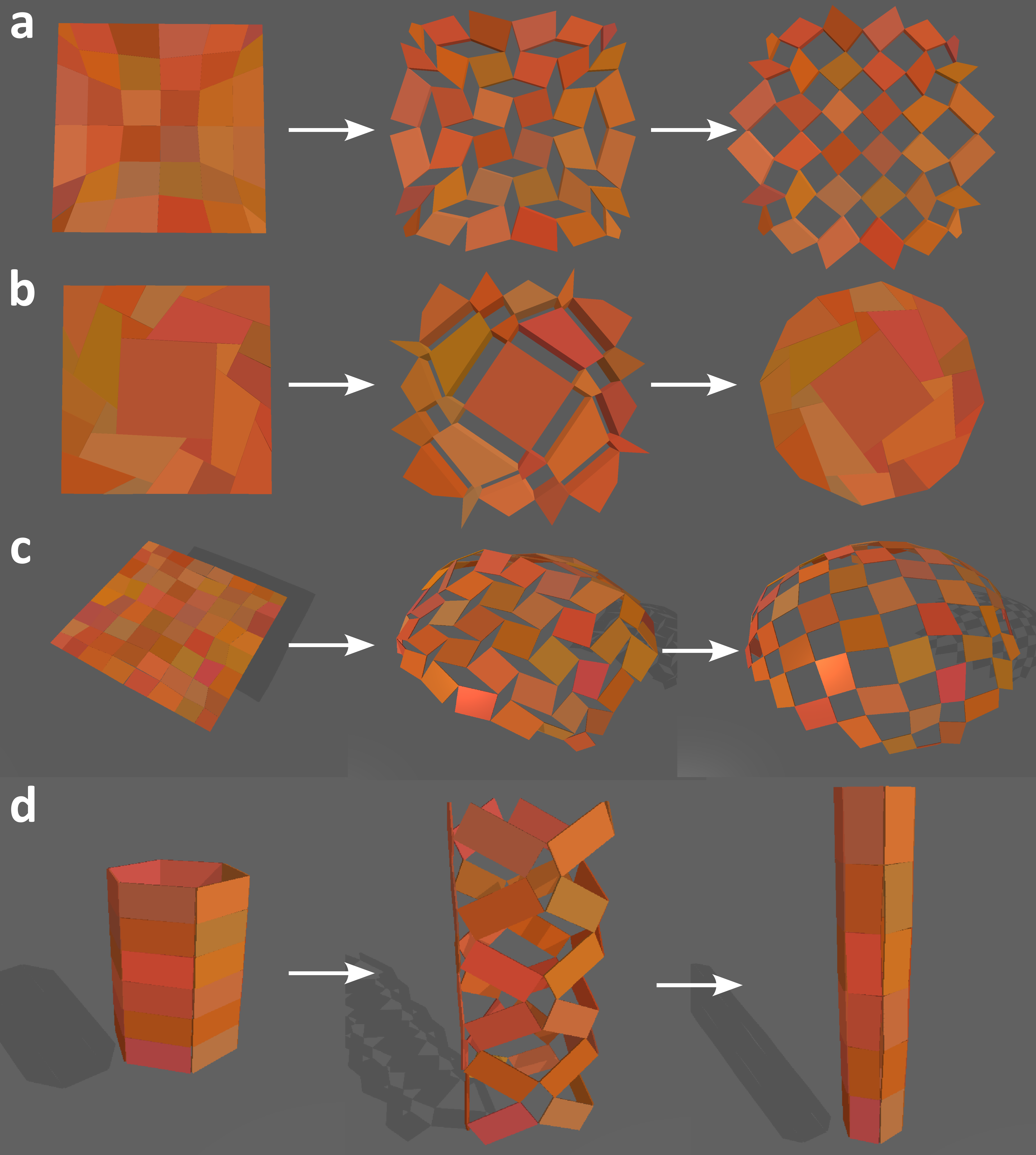}
    \caption{\textbf{Examples of 2D and 3D kirigami deployment achieved by PyKirigami.} \textbf{a}, The 2D deployment of the square-to-circle kirigami structure achieved by PyKirigami with automated radial expansion, stabilized by environmental forces. \textbf{b}, The 2D deployment of the compact reconfigurable kirigami model produced by PyKirigami. \textbf{c}, The 2D-to-3D deployment of the square-to-spherical-cap kirigami model achieved by PyKirigami, driven by the target-based force algorithm. \textbf{d}, The 3D-to-3D deployment and reconfiguration of the cylinder kirigami model achieved by PyKirigami, showcasing volumetric shape control.}
    \label{fig:results_combined}
\end{figure}

First, we consider the 2D expansion of the square-to-circle kirigami structure from~\cite{choi2019programming} as shown in Fig.~\ref{fig:results_combined}\textbf{a}. This example highlights the use of the basic Center-of-Mass Expansion Forces for straightforward radial deployment. More importantly, it demonstrates the essential role of Environmental and Stabilization Forces. By enabling a ground plane (Contact Modeling) and a downward gravity vector (Gravitational Loading), the deployment is constrained to the XY-plane, effectively simulating deployment on a physical surface and preventing the out-of-plane buckling that would otherwise occur in such a flexible structure.

Second, we showcase a compact reconfigurable kirigami model from~\cite{dudte2023additive}, which has multiple closed and compact contracted states (see Fig.~\ref{fig:results_combined}\textbf{b}). 
Unlike the open deployment in panel~\textbf{a}, this transition involves strong hinge rotations, contact, and friction, and is sensitive to dissipation and control. We therefore analyze its passive dynamics (implicit hinge elasticity) and show how to achieve stable, deterministic deployment via target-based actuation in the next subsection.

\subsubsection{3D Deployments and Volumetric Transformations}
The transformation from flat sheets to 3D surfaces, or between different 3D configurations, represents a significant challenge in kirigami simulation. Our next examples demonstrate how PyKirigami addresses these complex, out-of-plane dynamics.

The 2D-to-3D deployment of a square-to-spherical-cap kirigami model shown in Fig.~\ref{fig:results_combined}\textbf{c} exemplifies the necessity of the Target-Based Deployment Forces (Algorithm~\ref{alg:target_forces}). Simple expansion or manual interaction is insufficient to control the complex, coupled bending and stretching deformations required for this transformation. The vertex-based target model provides the precise control needed to guide the structure through its out-of-plane path, ensuring stable convergence to the desired spherical geometry. This case also underscores the importance of parameter tuning, where the actuation spring stiffness ($k_f$) and damping ($\mu_v$) are calibrated to balance deployment speed with stability.

Finally, we demonstrate a 3D-to-3D deployment with a compact reconfigurable cylinder kirigami structure as shown in Fig.~\ref{fig:results_combined}\textbf{d}. This highlights the ability of the Target-Based Deployment Forces for guiding complex, non-radial paths in three dimensions. Actuating the structure from a compact, folded state to its fully deployed form requires targets that define not just a final shape, but a collision-free volumetric transformation. This proves the model's suitability for simulating advanced systems like deployable space structures or reconfigurable robotics.

For completeness, while all examples in Fig.~\ref{fig:results_combined} use automated actuation (radial or target-based forces applied each simulation step), PyKirigami also supports a no-force mode and fully interactive runs in the GUI (manual actuation, pin/unpin, pause/reset). In Appendix~\ref{appendix:examples}, we show how to apply these functionalities to achieve the target shape of Fig.~\ref{fig:results_combined}\textbf{b}.

Besides the above examples, in Appendix~\ref{appendix:examples} we further discuss how more complex deployments of kirigami
structures can be achieved in PyKirigami. In particular, we demonstrate the ability of PyKirigami for handling the deployment of non-rigidly deployable kirigami structures as well as kirigami structures with highly complex deployment paths and complex topologies.

\subsection{Geometric Fidelity of Deployment}\label{subsec:geo-fidelity}
In PyKirigami, inter-tile joints are enforced iteratively by the underlying physics engine at each simulation timestep. Unlike reduced-coordinate formulations that encode kinematic loops as exact algebraic constraints, iterative enforcement inherently permits small constraint violation - a positional drift between nominally coincident vertices - whose magnitude depends on the number of solver iterations. To rigorously assess the impact of this numerical relaxation on kinematic accuracy, we perform a quantitative validation on two distinct kirigami structures: a parametric rectangular tessellation kirigami structure and a complex square-to-disk kirigami structure. In both cases, the simulated deployment trajectories are compared directly against exact analytical solutions derived from the geometric design parameters.

\subsubsection{Benchmark Case 1: Rectangular Tessellation}
We first consider a standard $M\times N$ rectangular tessellation kirigami structure with deployment angle $\theta$ as a benchmark. The theoretical vertex position $p^{\text{th}}_{i}$ at $\theta$ can be derived easily (see Appendix~\ref{appendix:analytic}). We initialize a $4\times 4$ grid in PyKirigami and actuate it from $\theta = 0$ to $\theta = \pi/2$.
At each time step, we compute the Euclidean error $\|p^{\text{sim}}_{i} - p^{\text{th}}_{i}\|$ for all vertices $i$. 

\begin{figure}[t!]
    \centering
    \includegraphics[width=\linewidth]{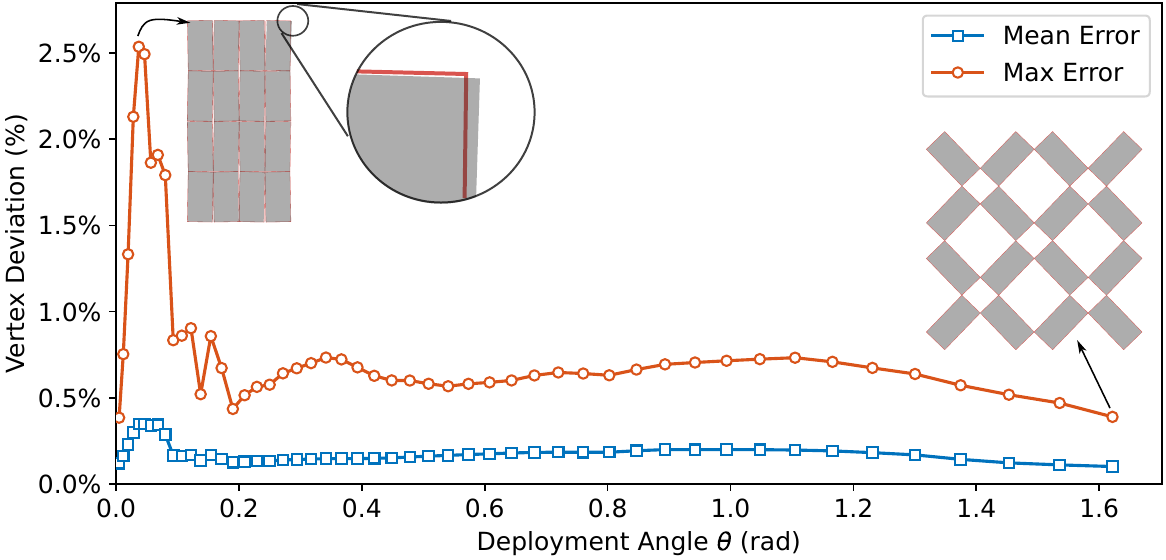}
    
    \caption{\textbf{Quantitative validation of PyKirigami using the rectangular tessellation kirigami structure.} The plot shows the maximal and mean vertex error versus the deployment angle $\theta$. Insets show a comparison of the simulated geometry (gray solid) against the theoretical trajectory (red wireframe) at two points. The left inset shows tessellation at the maximum deviation ($\theta\approx 0.04$) with the magnified view highlighting the drift. The right inset displays the fully deployed state, in which the error is negligible.}
    \label{fig:benchmark_tesse}
\end{figure}

As shown in Fig.~\ref{fig:benchmark_tesse}, the impulse-based solver possesses a very low positional error throughout the deployment process. The simulation initializes with a transient spike in maximum deviation $2.5\%$ as the solver resolves the ill-conditioned constraints near the compact state. Crucially, the mean relative error remains exceptionally low throughout the entire deployment process. This confirms that the transient spike is isolated to a few specific tiles and joints and does not propagate through the system.
The negligible mean error demonstrates that PyKirigami maintains vertex-level coherence, verifying the implementation of the topology constraint.

\subsubsection{Benchmark Case 2: Square-to-disk Structure}
\begin{figure}[t!]
    \centering
    \includegraphics[width=\linewidth]{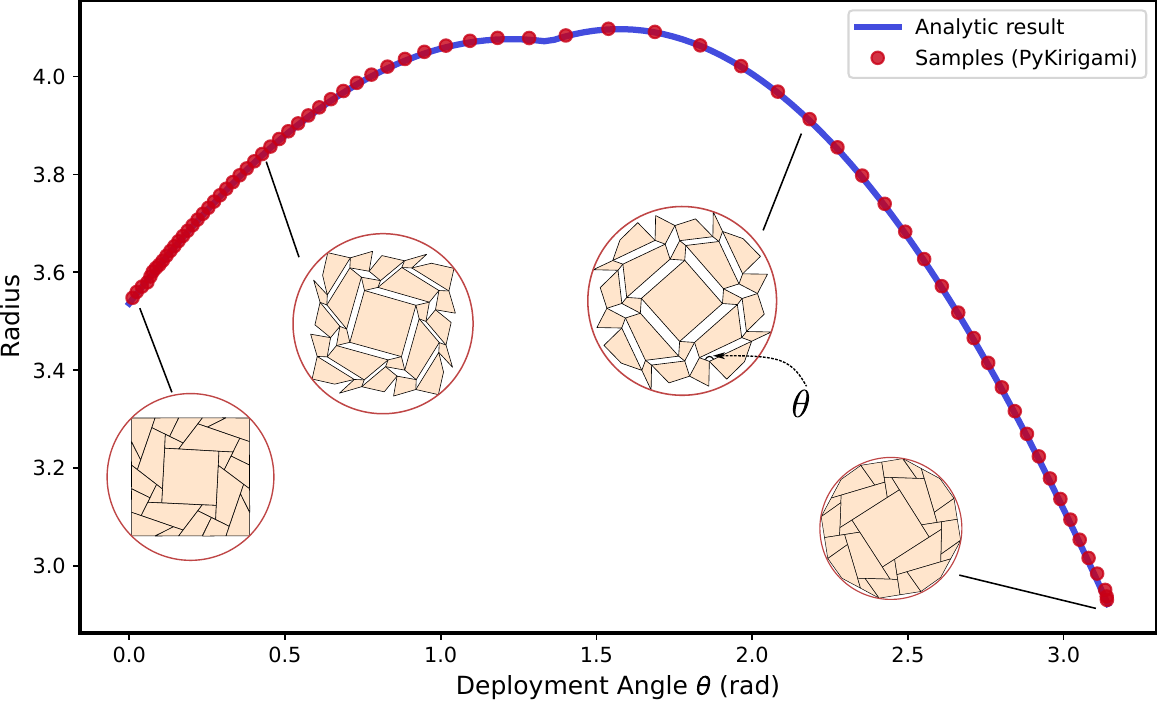}
    \caption{\textbf{Quantitative validation of PyKirigami using the square-to-disk kirigami structure.} Here, we plot the global radius $R$ versus the deployment angle $\theta$ for both the simulations obtained by PyKirigami (red dots) and the analytical result by~\cite{dudte2023additive} (blue solid line). The PyKirigami simulation data align closely with the analytical prediction, confirming that the solver captures the correct kinematic mode. The insets show four snapshots from the simulation.}
    \label{fig:benchmark_sq2disk}
\end{figure}

Having validated the solver's precision on the above benchmark case, we consider analyzing the deployment of the more complex square-to-disk structure as shown in Fig.~\ref{fig:results_combined}\textbf{b}. As explained in the theoretical construction of the structure in~\cite{dudte2023additive}, the deployment of it is also a single-degree-of-freedom mechanism (see Appendix~\ref{appendix:analytic}) where the global radius $R$ is a deterministic function of the tile rotation angle $\theta$. Therefore, in Fig.~\ref{fig:benchmark_sq2disk} we plot the global radius against the local deployment angle obtained by Pykirigami and compare the result with the analytical result in~\cite{dudte2023additive}, from which we can see that the discrete data samples obtained by PyKirigami at different time points of the deployment (red circles) show excellent agreement with the theoretical curve $R(\theta)$ (blue solid line). While micro-deviations exist at the hinge level (drift), the global topology remains locked to the correct deployment path, validating the tool for macroscopic shape analysis.

\subsection{Sensitivity Analysis and Parameter Tuning}
\label{subsec:sensitivity}
To validate the force model in Algorithm~\ref{alg:target_forces} and guide parameter selection for users of PyKirigami, we evaluate the influence of three key parameters: linear spring stiffness $k_f$, force damping $\mu_v$, and brick thickness $h$. Two representative models are considered, namely the square-to-disk model (focusing on 2D deployment) and the square-to-spherical-cap model (focusing on 3D deployment).  Deployment quality is measured by the maximum vertex error, i.e.\ the largest Euclidean distance between any simulated vertex and its target position, monitored throughout the simulation. Unless otherwise stated, all parameters take the default values listed in Table~\ref{tab:si_cli_params}.

\begin{figure}[!t]
    \centering
    \includegraphics[width=0.95\linewidth]{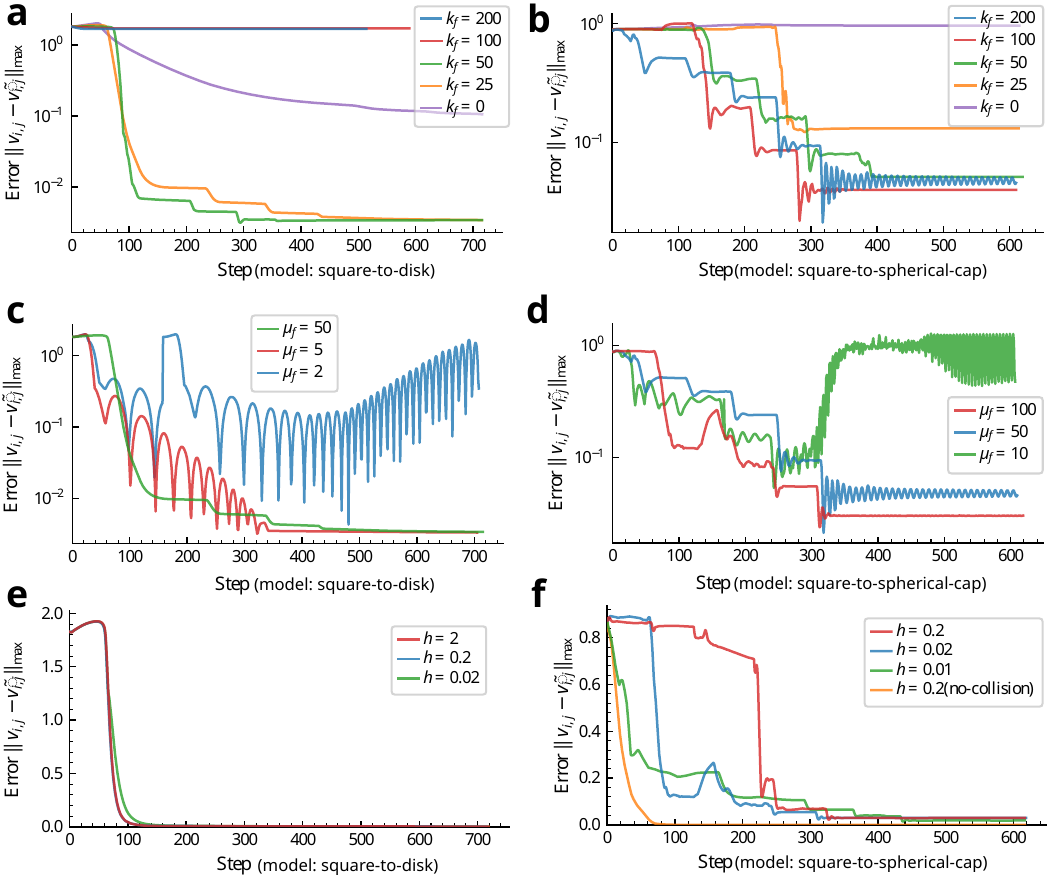}
    \caption{\textbf{Sensitivity of deployment quality to key simulation parameters on two representative models.} The left column (\textbf{a,c,e}) used the square-to-disk model with default thickness $h=0.2$; the right column(\textbf{b,d,f}) used the square-to-spherical-cap model with default thickness $h=0.02$. In each row, one parameter is varied. \textbf{a--b}, Linear spring stiffness $k_f \in \{0, 25, 50, 100, 200\}$ at fixed $k_\tau = 100$.  \textbf{c--d}, force damping $\mu_f$, swept at the respective stiffness $k_f=25$ and $k_f=200$. \textbf{e--f}, Brick thickness $h$.}
    \label{fig:sensitivity}
\end{figure}

\paragraph{Spring stiffness}
The stiffnesses $k_f$ and $k_\tau$ govern the magnitude of the translational and torsional actuation forces respectively. Sweeping $k_f \in \{0, 25, 50, 100, 200\}$ at fixed $k_\tau = 100$ reveals a qualitative difference between 2D and 3D deployments (Fig.~\ref{fig:sensitivity}\textbf{a,b}).

For the planar square-to-disk model (Fig.~\ref{fig:sensitivity}\textbf{a}), the dominant deployment motion is in-plane rotation of each tile about its geometric center. The linear spring force $k_f(\tilde{v}_{i,j}-v_{i,j})$ pulls each vertex directly toward its target along a chord, whereas the physical deployment path traces a rotational arc. Therefore, linear spring stiffness competes with required rotation rather than assisting it in the beginning, and the structure stalls at an intermediate state with the default linear spring stiffness $k_f=100$. The torsional-only case ($k_f=0$) avoids this issue, since $k_\tau$ generates torque directly aligned with the rotational degree of freedom. However, without any translational force, the structure would stall at a residual error of $\mathcal{O}(10^{-1})$ due to the existence of ground friction and damping effects. A moderate linear spring $k_f\in \{25,50\}$ balances both effects: it is small enough not to cancel out the required rotation in the beginning phase while still providing positional correction to drive the final error below $\mathcal{O}(10^{-2})$.

For the 2D-to-3D spherical-cap deployment(Fig.~\ref{fig:sensitivity}\textbf{b}), torsional actuation along $k_f = 0$ cannot drive tiles off the initial plane, so an extra linear spring $k_f$ is necessary. At $k_f=200$, visible oscillation appears. As discussed below, this reflects a damping mismatch. In both subplots, the staircase pattern ``a plateau followed by an apparent error drop'' is the signature of the adaptive stiffness mechanism: whenever deployments stall and maximal error is below the given tolerance, $k_f$ and $k_\tau$ are doubled until convergence resumes.

\paragraph{Translation damping}
Insufficient damping may cause oscillation or divergence when the spring force is large relative to the tile's inertia.  Using the stiffness values identified above ($k_f = 25$ for the square-to-disk; $k_f = 200$ for the spherical cap), we sweep $\mu_v$ on each model (Fig.~\ref{fig:sensitivity}\textbf{c,d}).  For the square-to-disk (Fig.~\ref{fig:sensitivity}\textbf{c}), reducing $\mu_v$ from 50 to 5 introduces oscillation near stiffness-doubling events but ultimately converges; $\mu_v = 2$ produces uncontrolled fluctuation from which the structure cannot recover.  For the spherical cap (Fig.~\ref{fig:sensitivity}\textbf{d}), the required damping is correspondingly higher: $\mu_v = 50$ already fluctuates at $k_f = 200$, which is stabilized at $\mu_v = 100$, while $\mu_v = 10$ leads to divergence.  These results confirm the expected coupling between $k_f$ and $\mu_f$: as stiffness increases, the required damping must increase proportionally to maintain a stable regime.

\paragraph{Thickness} Tile thickness $h$ affects deployment through the physical extent of the collision geometry. For the purely planar rigidly deployable model like square-to-disk(Fig.~\ref{fig:sensitivity}\textbf{e}), sweeping $h \in \{0.02, 0.2, 2\}$ produces nearly identical error curves. For the 2D-to-3D deployment (Fig.~\ref{fig:sensitivity}\textbf{f}), thickness becomes a controlling factor. Due to the non-rigid property of the initial planar pattern, the structure must first buckle with non-planar voids before converging to the target (see Appendix~\ref{appendix:complex}). The staircase plateaus in the plot correspond to this waiting period: the linear spring and torsional spring accumulate an imbalanced force distribution across the tile network, exploiting the kirigami structure to push some tiles upward and others downward until the collision geometry permits separation. The same pattern with larger thickness needs a spring with larger stiffness to be deployed. In contrast, the deployment with thickness $h=0.2$ when penetration between tiles is allowed is smooth, yielding a monotonically smooth convergence curve.

\subsection{Comparison with prior kirigami deployment simulator}

It is natural to compare our method with other open-source software for kirigami deployment simulation. Note that the prior 2D kirigami deployment simulator in~\cite{liu2022quasicrystal} utilizes the 2D rigid body physics engine Pymunk, focusing exclusively on 2D kirigami deployment. Also, the simulator only supports 2D manual interactive deployments and automatic radial deployments. By contrast, PyKirigami simulates both 2D and 3D kirigami deployment using the 3D PyBullet engine, and supports both 3D manual interactive deployments and more flexible target-based automatic deployments. Therefore, to provide a systematic and quantitative comparison, here we can only focus on a common 2D deployment task using the standard rectangular tessellations that both simulators can handle and compare their performance. Also, because of the inherent differences between the two engines (e.g., PyBullet's Projected Gauss--Seidel solver vs. Pymunk's sequential impulse solver), one cannot synchronize all solver-specific parameters that control fundamentally different internal mechanisms, such as the Error Reduction Parameter (ERP) and \texttt{error\_bias}, without introducing bias. Therefore, we focus on evaluating (1)~the achievable accuracy of each engine against a shared analytical ground truth, and (2)~the runtime scaling as a function of system size under matched kinematic conditions.

\begin{figure}
    \centering
    \includegraphics[width=0.95\linewidth]{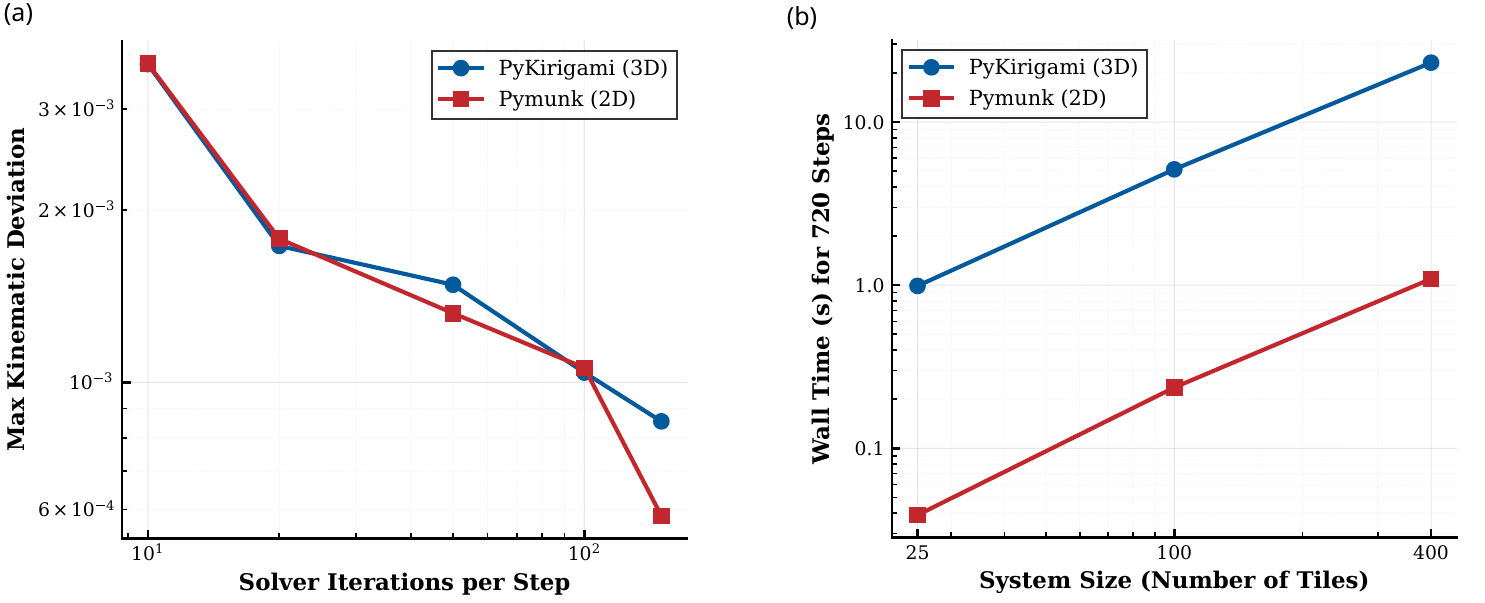}
    \caption{\textbf{Quantitative comparison between PyKirigami and Pymunk.} \textbf{(a)} Convergence of maximum kinematic deviation as a function of constraint solver iterations for a $10 \times 10$ grid where each grid is a unit square. Both engines exhibit monotonic decay, successfully reaching high-fidelity accuracy ($\mathcal{O}(10^{-4})$). \textbf{(b)} Runtime scaling as a function of system size over a 720-step deployment at a fixed 100 solver iterations. Both simulators exhibit sub-quadratic $\mathcal{O}(N^{1.2})$ algorithmic scaling. The constant vertical offset reflects the baseline computational cost of PyKirigami's 3D rigid-body formulation (6 DOF) compared to Pymunk's 2D planar formulation (3 DOF).}
    \label{fig:engine_comparison}
\end{figure}

We simulate the deployment of a 2D rectangular tessellation in both PyKirigami and Pymunk over 720 steps ($\Delta t = 1/240$\,s). To isolate the constraint solver's performance, gravity, friction, and damping are disabled, and collisions are bypassed. Fig.~\ref{fig:engine_comparison}\textbf{a} illustrates the convergence behavior for a $10 \times 10$ grid. We evaluate the maximum kinematic deviation across all joints, as this represents the worst-case tolerance failure in mechanism design. As solver iterations increase, both engines exhibit a strict monotonic decay in maximum deviation, demonstrating that both architectures can achieve high-fidelity deployments ($\mathcal{O}(10^{-4})$ deviation) given sufficient computational budget. 

To evaluate runtime scaling, we fix the solver iterations at 100 and measure the wall-clock time across varying grid resolutions (Fig.~\ref{fig:engine_comparison}\textbf{b}). For running a complete 720-step deployment, PyKirigami typically takes only a few seconds for moderately sized patterns and also only takes slightly over 10 seconds for a large pattern with $20\times 20 = 400$ tiles. In terms of comparison with Pymunk, note that both simulators exhibit highly efficient, sub-quadratic algorithmic scaling ($\sim\mathcal{O}(N^{1.2})$); a $16\times$ increase in system size (from $5\times5$ to $20\times20$) results in a $23\times$ increase in compute time for PyKirigami, and a $28\times$ increase for Pymunk. In absolute time, one may see that Pymunk is faster than PyKirigami for equivalent grid sizes. This difference is fundamentally architectural: Pymunk solves strictly 2D planar dynamics (3 degrees of freedom per body), whereas PyKirigami utilizes a maximal-coordinate 3D formulation (6 degrees of freedom per body). This baseline computational overhead is the necessary and justified cost of PyKirigami's 3D architecture, which natively captures phenomena that 2D tools fundamentally cannot simulate, such as tile thickness, out-of-plane buckling, and complex 3D self-collisions.

\section{Discussion}
In this work, we have developed PyKirigami, an efficient and open-source computational tool for the deployment simulation of kirigami structures. Specifically, we have demonstrated the ability of our PyKirigami simulator for handling different 2D and 3D kirigami structures with different customized settings in terms of the tile geometries, tile connections, and other deployment parameters. Altogether, the significantly expanded functionalities of PyKirigami over prior open-source kirigami simulation software facilitate the computational modelling process for kirigami structures, thereby paving a new way for the design of shape-morphing metamaterials. 

While PyKirigami provides an useful framework for rigid-body dynamics, we acknowledge its current limitations. First, note that the simulator considers the 3D kirigami tiles and their connections without capturing material elasticity, plasticity, or fracture. Also, high-fidelity stress/strain analysis would still require FEA methods. Besides, the current implementation of the simulator assumes the preservation of the kirigami tile connections throughout the deployment, while real-time cutting or topology changes are not supported. Our simulator is therefore best positioned as a tool for rapid prototyping, kinematic analysis, and exploring the global deployment dynamics of complex kirigami assemblies. For instance, it may be utilized for rapidly simulating the deployment effects and trajectories of kirigami structures under large-scale parametric sweeps across different deployment ranges, collision constraints, and morphing configurations, thereby forming a dedicated kinematic pre-screening pipeline for kirigami design.

In the future, we plan to enhance the functionalities of PyKirigami to include the consideration of elasticity and contact-induced deformations. Specifically, besides rigid-body dynamics, the underlying PyBullet engine also supports certain soft-body and deformable-object simulations, though they are known to be prone to numerical instability under large timesteps and body collisions. By carefully adjusting the relevant parameter setups, it may be possible to integrate these simulations into PyKirigami to allow for a wider range of kirigami deployment effects. Another future direction is to develop PyKirigami into an interactive website similar to Origami Simulator~\cite{origamisimulator}, making the software tool more user-friendly and accessible.

\noindent \textbf{Data Availability} \  The code and data are available on GitHub at \url{https://github.com/andy-qhjiang/PyKirigami}.\\

\bibliographystyle{elsarticle-num}
\bibliography{reference_new}

\clearpage

\appendix

\begin{center}
    \Large{\textbf{Appendix}}
\end{center}

\section{Overview} \label{appendix:overview}
PyKirigami is an open-source Python toolbox for simulating the 2D and 3D deployment of general kirigami structures. The simulator utilizes the PyBullet physics engine and is capable of simulating 2D-to-2D, 2D-to-3D, and 3D-to-3D kirigami deployments with interactive controls. The toolbox is freely available on GitHub:
\begin{center}
    \url{https://github.com/andy-qhjiang/PyKirigami}
\end{center}

The detailed installation instructions, usages, and example projects can be found in the user manual: 
\begin{center}
    \url{https://github.com/andy-qhjiang/PyKirigami/wiki}
\end{center}

In this supplementary document, we focus on how we realize the key features and functionalities of PyKirigami to make it easy to be understood and extended by interested users.

\section{Implementation Walkthrough} \label{appendix:implementation}

\subsection{Coordinate System Management}
A fundamental concept in PyBullet, and 3D simulations in general, is the distinction between \textbf{world coordinates} and \textbf{local coordinates}. Understanding this is key to efficiently defining connections and analyzing the kirigami structure's motion.

\paragraph{Frames and notation}
We first define the following:
\begin{itemize}
    \item World frame: The fixed global frame.
    \item Local (base) frames: one per tile, with origins at the tile centers $O_1$ (tile A) and $O_2$ (tile B).
    \item Rotations: $R_1,R_2\in\mathrm{SO}(3)$ map local coordinates to world for tiles A and B.
    \item Coordinates: $\mathbf{p}_l$ and $\mathbf{q}_l$ are the local coordinates of $P$ and $Q$; $\mathbf{p}_w$ and $\mathbf{q}_w$ are the corresponding world coordinates.
\end{itemize}

\paragraph{Local-world transforms}
The exact mappings are
\begin{equation}
\mathbf{p}_w = R_1\,\mathbf{p}_l + O_1,\qquad
\mathbf{q}_w = R_2\,\mathbf{q}_l + O_2,
\end{equation}
and, inversely,
\begin{equation}
\mathbf{p}_l = R_1^{\top}\,(\mathbf{p}_w - O_1),\qquad
\mathbf{q}_l = R_2^{\top}\,(\mathbf{q}_w - O_2).
\end{equation}

\begin{figure}[t]
    \centering
    \includegraphics[width=\linewidth]{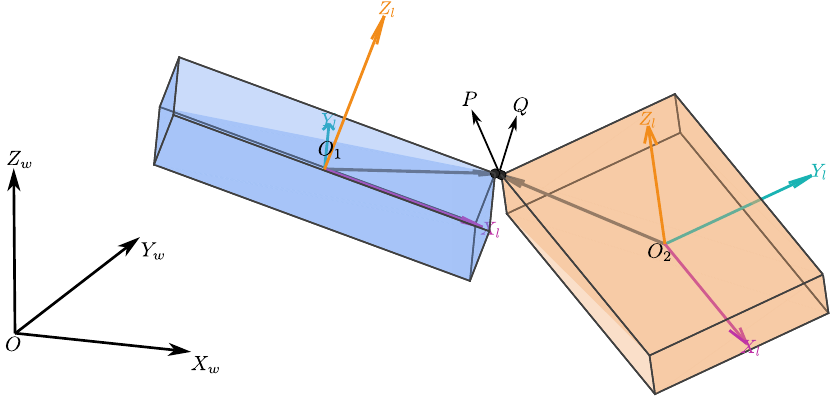}
    \caption{\textbf{Coordinate systems in PyBullet.} The fixed global axes $(X_w, Y_w, Z_w)$ define the world coordinate system. Each kirigami tile (represented by a box) has its own local coordinate system $(X_l, Y_l, Z_l)$ attached to its center. A corner vertex of the blue box, such as $P$, has constant local coordinates $\mathbf{p}_{l}=P-O_{1}$ in its body's frame, but its world coordinates $\mathbf{p}_{w}$ change as the body moves.}
    \label{fig:coordinates}
\end{figure}

See Fig.~\ref{fig:coordinates} for an illustration. For convenience, we provide helper functions that implement the local-world transforms:

\begin{python}
import pybullet as p
import numpy as np

def local_to_world(body_id, p_local):
    pos, orn = p.getBasePositionAndOrientation(body_id)  
    # pos is the world frame position of tile center O_i
    R = np.array(p.getMatrixFromQuaternion(orn)).reshape(3, 3)  # R_i
    return (R @ np.asarray(p_local)) + np.asarray(pos)

def world_to_local(body_id, p_world):
    pos, orn = p.getBasePositionAndOrientation(body_id)
    R = np.array(p.getMatrixFromQuaternion(orn)).reshape(3, 3)
    return (R.T @ (np.asarray(p_world) - np.asarray(pos)))
\end{python}

\subsection{Object Creation in PyBullet}
Given the planar vertices of a tile in world coordinates, we construct a 3D ``brick'' of thickness $t$. The core of this process is to create two distinct geometric representations for the tile, both defined in its local (base) frame: a simple \textbf{collision shape} for physics calculations and a detailed \textbf{visual shape} for rendering. This separation optimizes performance while ensuring visual fidelity.

\begin{figure}[t]
    \centering
    \includegraphics[width=\linewidth]{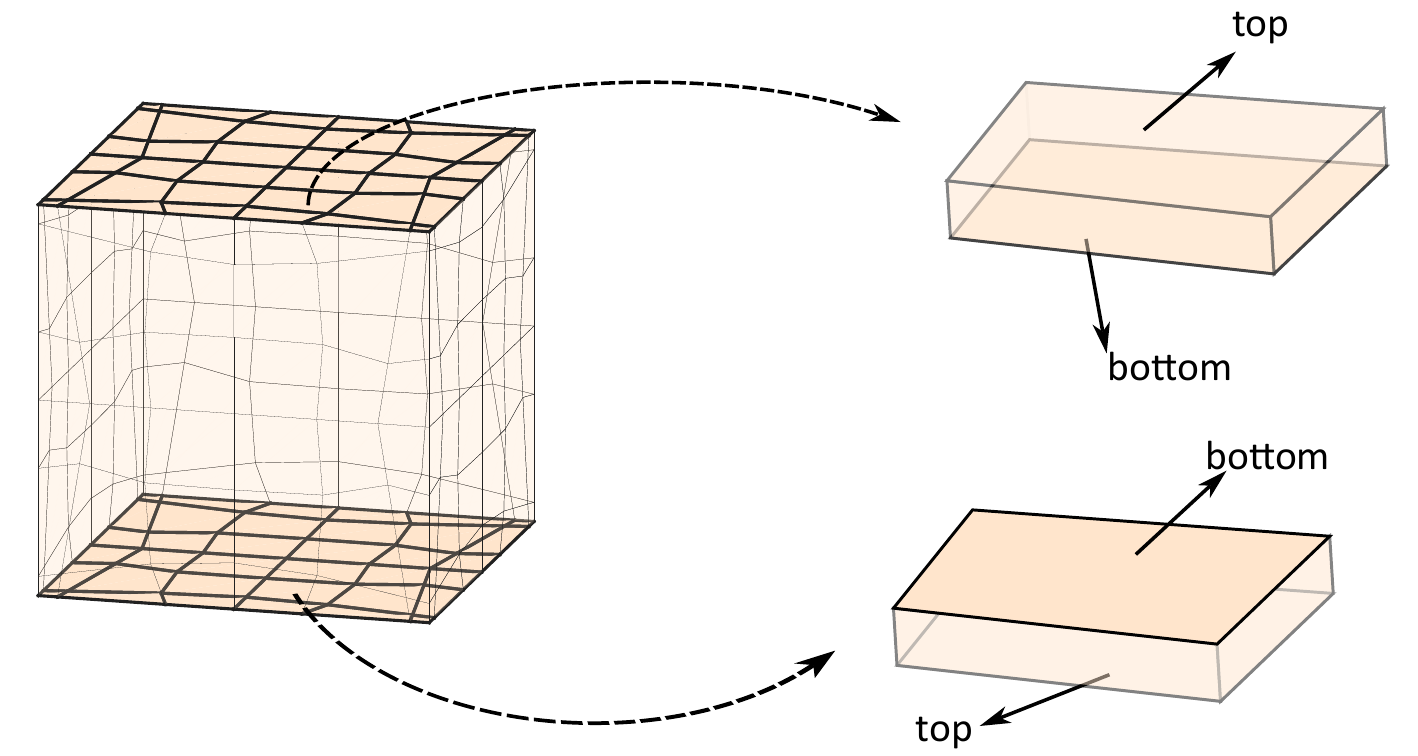}
    \caption{\textbf{An illustration of face extrusion for 3D kirigami structures.} Here, note that the quadrilateral patches on opposite faces of the cube are congruent, but their orientations are different. As the outward normal flips ($\mathbf{n}\mapsto -\mathbf{n}$), the vertex order must be reversed to maintain outward-facing normals. If the top face uses $(v_0,v_1,v_2,v_3)$ in counter-clockwise order (viewed from outside), the corresponding bottom face should be indexed $(v_0,v_3,v_2,v_1)$.}
    \label{fig:normal_direction}
\end{figure}

\paragraph{Extrusion and face orientation}
As described in the main text (see also main text Fig.~2), we define the tile's outward unit normal $\mathbf{n}$ in world coordinates and extrude along $+\mathbf{n}$ to obtain the top face, while the original tile becomes the bottom face. In Fig.~\ref{fig:normal_direction}, we further illustrate this idea for 3D kirigami structures. Specifically, this local-normal convention is essential in 3D-to-3D deployments, where the world $z$-axis generally does not align with the tile’s normal.

\paragraph{Collision Shape Construction}
The collision shape is used by the physics engine to detect contact, resolve forces, and simulate dynamics. For stability and speed, we set it as a simple convex volume since this already meets our requirements for the simulation. After extrusion, we pass the list of local coordinates
$$
\{u_j = v_j - O, u_j^t = v_j^t - O\}
$$
to \texttt{p.createCollisionShape}. PyBullet automatically computes the convex hull of the vertices to form an efficient, though invisible, collision mesh (Fig.~\ref{fig:brick_creation}). Below is an example of the collision shape creation for a triangular tile:

\begin{figure}[t]
\centering
\includegraphics[width=1.0\linewidth]{F10_SI_brick_creation_demo.pdf}
\caption{\textbf{Brick creation in the local frame.} \textbf{a}, Extrusion: We compute the world-space top vertices $v_j^t = v_j + t\cdot\mathbf{n}$ and the base origin $O$ (tile center), then convert both bottom and top vertices to local coordinates via $u_j = v_j - O$ and $u_j^t = v_j^t - O$. \textbf{b}, Mesh construction: We build a convex collision vertex cloud (local) and a per-vertex-normal visual mesh (local) with cap and side faces.}
\label{fig:brick_creation}
\end{figure}

\begin{python}
    # PyBullet computes the convex hull based on local coordinates
    # which is exactly the triangular prism
    col_shape = p.createCollisionShape(
        shapeType=p.GEOM_MESH,
        vertices=[
                [x1, y1, z1], #  coordinates of u0
                [x2, y2, z2], #  coordinates of u1
                [x3, y3, z3], #  coordinates of u2
                [x4, y4, z4], #  coordinates of u0t
                [x5, y5, z5], #  coordinates of u1t
                [x6, y6, z6], #  coordinates of u2t 
                ]
    )
\end{python}

\paragraph{Visual Shape Construction}
The visual shape provides a detailed mesh for rendering sharp edges and correct lighting. The key to achieving this is understanding that PyBullet's graphics API assigns normals on a \textbf{per-vertex} basis. To create flat-shaded faces with hard edges, we must often duplicate geometric vertices in our data buffers, pairing each copy with the correct face normal. We illustrate this by building a single rectangular side face of the brick illustrated in Fig.~\ref{fig:brick_creation}\textbf{b}, defined by local bottom vertices $(u_0, u_1)$ and top vertices $(u_0^t, u_1^t)$. Below is an example:

\begin{python}
# Define the 4 unique vertices of the rectangular face
vis_vertices = [
    u0,  # index 0
    u1,  # index 1
    u1t, # index 2
    u0t  # index 3
]

# The API requires one normal per vertex. Identical For a flat face.
n_side = calculate_face_normal(v0, v1, v1t)
vis_normals = [n_side]*len(vis_vertices)

# Define two triangles using indices into the vertex/normal lists
vis_indices = [
    0, 1, 2,  # First triangle: (v0, v1, v1t)
    0, 2, 3   # Second triangle: (v0, v1t, v0t)
]

# These lists are passed to create the visual component of the body
vis_shape = p.createVisualShape(
    shapeType=p.GEOM_MESH,
    vertices=vis_vertices,
    indices=vis_indices,
    normals=vis_normals,
    rgbaColor=[...],
    specularColor=[0.8, 0.8, 0.8] # Glare; smaller is less shiny
)
\end{python}

\paragraph{Visual customization}
The tile's appearance can be modified at creation time or dynamically via the \texttt{rgbaColor} parameter. We use sky blue color as an indicator of fixing an object during simulation:
\begin{python}
p.changeVisualShape(bodyUniqueID=body_id, rgbaColor=[0.53, 0.81, 0.92, 1.0])
\end{python}

\vspace{-2mm}

\paragraph{Assembly}
With both shapes defined in the local frame, instantiating the rigid body is straightforward. The only world quantity to be supplied is the base pose $(O,R)$; typically $R=I$ at load time.

\begin{python}
def create_brick_body(col_shape, vis_shape, base_origin_world, mass=1.0, base_quat=[0,0,0,1]):
    return p.createMultiBody(
                baseMass=mass,
                baseCollisionShapeIndex=col_shape,
                baseVisualShapeIndex=vis_shape,
                basePosition=base_origin_world, # world position of local origin
                baseOrientation=base_quat # usually identity; optional
)
\end{python}

This completes the brick creation pipeline: (i) extrude and define local geometry, (ii) build a convex collision shape and a flat-shaded visual mesh with per-vertex normals via vertex duplication, and (iii) instantiate the rigid body at its world pose.

\subsection{Creating Inter-Tile Constraints}\label{subsec:constraint}

With each tile defined as a rigid body in its own local frame, we now connect them to form the kirigami structure. PyKirigami uses PyBullet's constraint system to model the joints between tiles. All constraints are defined by specifying pivot points in the \textbf{local coordinates} of the connected bodies, leveraging the framework established in the previous sections.

\paragraph{Constraint Validation}
Before creating any physics objects, it is crucial to validate that the provided vertex and constraint data are consistent. For a kirigami structure to be physically valid, the world-space positions of any two vertices designated as a pivot pair must coincide. PyKirigami performs this check before entering the simulation loop:

\begin{python}
def validate_constraints(vertices, constraints, tolerance=1e-6):
    # Unpack constraint data (e.g., for a spherical joint)
    for i, constraint in enumerate(constraints):
        f1, v1, f2, v2 = constraint[:4]
        # Get vertex positions for constraint endpoints
        p1_world = np.array(vertices[f1][3*v1:3*v1+3])
        p2_world = np.array(vertices[f2][3*v2:3*v2+3])
        # Check distance
        dist = np.linalg.norm(np.array(p1_world) - np.array(p2_world))
        if dist > tolerance:
            raise ValueError(f"Constraint validation failed: distance is {dist}")
\end{python}

\paragraph{Spherical Joint}
As illustrated in Fig.~\ref{fig:coordinates}, a connection is defined by a pair of pivot points, such as $P$ on tile A and $Q$ on tile B. We provide PyBullet with the constant local coordinates of these pivots, $\mathbf{p}_l$ and $\mathbf{q}_l$. The physics engine then ensures that the world-space positions of these points, $\mathbf{p}_w$ and $\mathbf{q}_w$, satisfy the constraint equation 
\begin{equation}
R_1\cdot\mathbf{p}_{l} + O_1 \;=\; R_2\cdot\mathbf{q}_{l} + O_2,
\end{equation}
at every time step. It is implemented with a single \texttt{JOINT\_POINT2POINT} constraint as follows:

\begin{python}
# create spherical joint and return cid (constraint ID)
def connect_spherical(bodyA, bodyB, pA_local, pB_local):
    return p.createConstraint(
                    parentBodyUniqueId=bodyA,
                    parentLinkIndex=-1,
                    childBodyUniqueId=bodyB, 
                    childLinkIndex=-1,
                    jointType=p.JOINT_POINT2POINT, 
                    jointAxis=[0,0,0],
                    parentFramePosition=pA_local,
                    childFramePosition=pB_local
)

\end{python}

Since the kirigami bricks have thickness, we must specify whether a pivot lies on the bottom face (original vertices) or the top face (extruded vertices) as illustrated in the main text. Our constraint file format supports an optional face specifier \texttt{t} (1 for bottom, 2 for top). When parsing, we select the appropriate local vertex vector, either $\mathbf{u}_j$ (bottom) or $\mathbf{u}_j^t$ (top), to use as the pivot.

\paragraph{Hinge Joint} A hinge restricts rotation to a single axis, providing one rotational degree of freedom. In PyKirigami, we implement a hinge by using \textbf{two} non-coincident spherical joints. The line connecting the two pivot pairs defines the hinge axis. This method is numerically stable and works well for bodies with finite thickness:

\begin{python}
def connect_hinge(bodyA, bodyB, pA1, pB1, pA2, pB2):
    cid1 = connect_spherical(bodyA, bodyB, pA1, pB1)
    cid2 = connect_spherical(bodyA, bodyB, pA2, pB2)
    return [cid1, cid2]
\end{python}

\paragraph{Fixed Joint}
This joint removes all six relative degrees of freedom between two bodies, effectively welding them together. We use the native \texttt{JOINT\_FIXED} constraint. This is particularly useful for anchoring a tile to the world (by constraining it to a static body) or for creating rigid composite structures:

\begin{python}
def fix_object_to_world(body_id):
    pos, orn = p.getBasePositionAndOrientation(body_id)
    constraint_id = p.createConstraint(
        parentBodyUniqueId=body_id,
        parentLinkIndex=-1,
        childBodyUniqueId=-1,  # World frame
        childLinkIndex=-1,
        jointType=p.JOINT_FIXED,
        jointAxis=[0, 0, 0],
        parentFramePosition=[0, 0, 0],
        childFramePosition=pos,  # tile center O_i in world
        parentFrameOrientation=[0, 0, 0, 1], 
        childFrameOrientation=orn    # tile orientation in world
    )
    p.changeConstraint(cid, maxForce=max_force)
    return cid
\end{python}

\textbf{Remark.} To make a body immovable, one could either use a \texttt{JOINT\_FIXED} constraint or set the body's mass to zero. For an articulated system like a kirigami structure, the constraint-based approach is recommended since an anchored tile must still be able to transmit forces and torques from the rest of the structure to the world. A fixed joint allows for this, as the body remains a dynamic participant in the simulation. In contrast, setting a body's mass to zero removes it from the dynamics calculation, which can cause solver instability when other dynamic bodies are joined to it. The practice of setting mass to zero is best reserved for defining permanent, static environment geometry like floors or immovable obstacles.

\paragraph{Automatic Connection-Type Detection}
When a target file is provided, our PyKirigami tool can automatically determine whether to connect the bottom or top face of each tile pair (controlled by the \texttt{--auto\_detect\_connections} flag). The decision is made by analyzing the target geometry:
\begin{itemize}
    \item \textbf{Shared-edge case.} If two tiles share an edge in the target, the signed dihedral angle $\delta$ is computed via $\delta = \operatorname{atan2}((\mathbf{n}_a \times \mathbf{n}_b)\cdot \mathbf{e},\; \mathbf{n}_a \cdot \mathbf{n}_b)$, where $\mathbf{n}_a,\mathbf{n}_b$ are face normals and $\mathbf{e}$ is the shared edge direction. $\delta < \pi$ indicates a ``valley fold'' (top connection); $\delta \ge \pi$ indicates a ``mountain fold'' (bottom connection).
    \item \textbf{Negative-space case.} If tiles meet only at a single vertex, the face normals are projected onto the inter-centroid vector. Converging normals indicate a ``valley fold'' (top connection); diverging normals indicate a ``mountain fold'' (bottom connection); neutral cases default to bottom.
    \item \textbf{Planar-to-planar case.} If both initial and target configurations are planar, both faces are connected (type~3) for stability.
\end{itemize}

\subsection{Interactive Control and Simulation State Management}

The runtime system comprises three layers:
\begin{itemize}
    \item Entry point and GUI loop (run\_sim.py)
    \item Simulation (initialization and force selection)
    \item Controllers: SimulationController (runtime stepping/state) and InteractionController (pin/unpin, snapshots, and save data)
\end{itemize}

\paragraph{Simulation: initialization and force selection.}
The Simulation class constructs all rigid bodies and constraints from files, configures damping/environment, and prepares data for runtime control. It returns a dictionary \texttt{simulation\_data} for downstream controllers, summarized in Table~\ref{tab:simdata}.

\begin{table}[h!]
\centering
\caption{\textbf{Contents of the \texttt{simulation\_data} dictionary passed to controllers.}}
\label{tab:simdata}
\begin{tabularx}{\linewidth}{@{}lX@{}}
\toprule
\textbf{Key} & \textbf{Description} \\
\midrule
\texttt{args} & Parsed command-line configuration (timestep, damping, thickness, modes, etc.). \\
\texttt{bricks} & List of PyBullet body IDs (one per tile). \\
\texttt{local\_coords} & Bottom-face local vertex coordinates per tile (used for local-to-world transforms during runtime). \\
\texttt{constraint\_ids} & List of PyBullet constraint IDs created between tiles. \\
\texttt{visual\_mesh} & List of visual mesh for each tile to be exported as an OBJ file. \\
\texttt{target\_vertices} & Optional world-space target vertices per tile (bottom face), used for target-based actuation. \\
\bottomrule
\end{tabularx}
\end{table}

\paragraph{SimulationController: runtime state and stepping.}
Two controllers drive the runtime behavior, each consuming the entries in \texttt{simulation\_data}:

\begin{itemize}
\item Stepping: When \texttt{step\_simulation()} is called, if the simulation is not paused, it first calls the \texttt{apply\_forces()} function (which selects the appropriate actuation model) and then calls \texttt{p.stepSimulation()}.
\item State Management: It handles \texttt{reset\_simulation()} by calling the function \texttt{initialize\_} 
\texttt{simulation()} to rebuild the world from scratch. 
\end{itemize}

\paragraph{InteractionController: user input and visual feedback.}
The InteractionController handles real-time user input from the mouse and provides visual feedback. It also uses \texttt{simulation\_data} to identify which objects are part of the kirigami structure.
\begin{itemize}
\item Mouse Events: In \texttt{process\_mouse\_events()}, it performs a ray cast from the camera through the mouse cursor. If the ray hits an object whose ID is in the bricks list from \texttt{simulation\_data}, it triggers the pin/unpin action.

\item Data Output: An OBJ file is exported.
\end{itemize}

\paragraph{Main loop} 
The entry point \texttt{run\_sim.py} wires everything together and runs an interactive loop. The core structure is given in Algorithm~\ref{alg:main_loop}:

\begin{algorithm}[H] 
\caption{Interactive main loop}
\begin{algorithmic}[1]
\STATE Initialize physics, Simulation, SimulationController, InteractionController
\WHILE{Bullet client is connected}
\STATE keys $\leftarrow$ getKeyboardEvents()
\FOR{each (key, handler) in handlers}
\IF{key was triggered}
\STATE call handler()
\IF{key == 'q'} \STATE break loop \ENDIF
\ENDIF
\ENDFOR
\STATE InteractionController.process\_mouse\_events()
\STATE SimulationController.step\_simulation()
\STATE sleep(timestep)
\ENDWHILE
\STATE disconnect()
\end{algorithmic}
\label{alg:main_loop}
\end{algorithm}

For completeness and reproducibility, we summarize the default interactive controls used in PyKirigami's GUI mode. These inputs allow manual actuation, dynamic boundary condition changes, camera manipulation, and on-demand export. Key bindings can be customized in the codebase; the mappings shown in Table~\ref{tab:controls} reflect the default configuration used in this paper.

\begin{table}[h!]
\centering
\caption{\textbf{Interactive Control Scheme for the PyKirigami Simulator.}}
\label{tab:controls}
\begin{tabularx}{\columnwidth}{lX} 
\toprule
\textbf{Control Input} & \textbf{Functionality} \\
\midrule
Left-click + drag    & Apply external, damped forces to tiles for manual guidance and actuation during deployment. \\
Right-click          & Toggle fixed-world constraints on a selected tile. \\
\texttt{Space bar}           & Toggle pause/resume the physics simulation.\\
\texttt{F}           & Toggle automatic deployment forces on/off.\\
\texttt{R}           & Reset the simulation to its initial configuration.\\
\texttt{O}           & Export the result as an OBJ file.\\
\texttt{Q}           & Quit the simulation application. \\
Scroll / Ctrl+drag   & Zoom, rotate, and pan the camera for detailed visual inspection. \\
\bottomrule
\end{tabularx}
\end{table}

\subsection{Physical Environment Construction: CLI Parameters and Defaults}
This section documents the primary command-line interface (CLI) flags used to configure and run the simulations. The table below groups related parameters and lists their default values and usage. For simplicity and portability, input files can be organized into a single directory within the \texttt{data/} folder and loaded using the \texttt{--model} flag. The CLI parameters and default values are given in Table~\ref{tab:si_cli_params}.

\begin{table}[h!]
\centering
\caption{\textbf{Command-Line Interface (CLI) parameters and default values.} Parameter groups are distinguished by alternating background colors.}
\label{tab:si_cli_params}
\begin{tabularx}{\linewidth}{@{} l l X @{}}
\toprule
\textbf{Flag} & \textbf{Default} & \textbf{Description} \\
\midrule
\multicolumn{3}{@{}l}{\textit{\textbf{Model and File Configuration}}} \\
\texttt{--model} & (none) & Specifies a model folder \texttt{data/<name>/} containing input files like \texttt{vertices.txt}, \texttt{constraints.txt} and optional \texttt{target.txt}.\\

\rowcolor{Gray} \multicolumn{3}{@{}l}{\textit{\textbf{Simulation Control and Stepping}}} \\
\rowcolor{Gray} \texttt{--timestep} & 1/240 & Discrete virtual time increment for solving rigid-body dynamic inside PyBullet \\

\multicolumn{3}{@{}l}{\textit{\textbf{Physical Environment}}} \\
\texttt{--gravity} & 0 & Gravitational acceleration in $-z$. \\
\texttt{--ground\_plane} & (switch) & Adds ground plane to the simulation. \\
\texttt{--filter\_collision} & (switch) & Cancel collision detection during simulation.\\

\rowcolor{Gray} \multicolumn{3}{@{}l}{\textit{\textbf{Actuation Model}}} \\
\rowcolor{Gray} \texttt{--cm\_expansion} & (switch) & Enables center-of-mass radial expansion force model. \\
\rowcolor{Gray} \texttt{--spring\_stiffness} & 100 & Linear spring stiffness $k_s$ for target-based forces. \\
\rowcolor{Gray} \texttt{--torque\_stiffness} & 100 & Torsional spring stiffness $k_\tau$ for target-based torques. \\
\rowcolor{Gray} \texttt{--force\_damping} & 50 & Actuator (viscous) damping $\mu_v$ for target-based forces. \\
\rowcolor{Gray} \texttt{--torque\_damping} & 2 & Actuator (viscous) damping $\mu_\omega$ for target-based torques. \\
\multicolumn{3}{@{}l}{\textit{\textbf{Geometry and Visualization}}} \\
\texttt{--brick\_thickness} & 0.02 & Thickness of extruded tiles (m). \\

\bottomrule
\end{tabularx}
\end{table}

\subsection{Self-Penetration Avoidance}  \label{appendix:self_penetration}
PyKirigami tries to avoid self-penetration during the deployment simulation through several complementary mechanisms, operating at different levels of the simulation pipeline:
\begin{enumerate}
    \item \textbf{Physical-engine-level collision detection (per time step).} The most fundamental layer of protection is provided by PyBullet's built-in rigid body collision detection, which runs at every simulation step. When two tiles approach within a configurable collision margin $\epsilon$, the physical engine applies a corrective impulse to prevent interpenetration. This ensures that, even if actuation forces attempt to drive two tiles into the same spatial region, the solver will resist penetration up to numerical tolerance. Notice that a small residual overlap (on the order of the collision margin $\epsilon$, typically $\sim 1\%$ of tile size) is unavoidable, which is standard in rigid-body simulation.
    \item \textbf{Automatic connection-type detection at the tile level.} Potential penetration between tiles exists due to the extrusion of planar polygons. For example, the boundary of the top face of both tiles would collide if both tiles are driven to a valley shape while they are connected by a bottom joint. Therefore, we provide users with an option for PyKirigami to automatically classify each inter-tile connection based on the target vertices, ensuring that the structure's connectivity topology is consistent with the actuation forces applied to each tile.
    \item \textbf{Kabsch alignment for global correspondence.} A more subtle source of self-intersection of kirigami patterns during automatic deployment arises from incorrect global target assignment. If the target vertex positions are naively matched to source tiles without accounting for global orientation, a tile on the left side of the pattern may be assigned a target on the right side, and vice versa. The resulting crossed actuation force would drive different parts of the structure through each other, causing catastrophic self-intersection that collision detection alone cannot prevent. PyKirigami resolves this by applying the Kabsch algorithm to compute the optimal rigid-body alignment between all points from vertices file and target file before applying target-based force. This can avoid crossing trajectories to some extent.

    \item \textbf{User-defined intermediate targets for complex deployment.}
    For highly complex target shape changes or deployment trajectories, we suggest decomposing the deployment into a sequence of simpler sub-deployments, analogous to waypoint-based motion in robotics. We have functions integrated in PyKirigami that export the current position of the structure and convert it to the same format as target vertices, so that the user can decompose the deployment into pieces by selecting the original pattern and target pattern themselves. 
\end{enumerate}

\section{Example Configurations and Simulation Details} \label{appendix:examples}

In the main text, we presented several 2D-to-2D, 2D-to-3D, and 3D-to-3D deployments obtained with PyKirigami. Here we provide additional snapshots and, crucially, the corresponding physical environment and solver settings used for each case so that readers can reproduce the results precisely. All CLI flags follow Table~\ref{tab:si_cli_params}. All simulations were run with a fixed timestep $\Delta t = 1/240$\,s and $n_{\text{sub}}=20$ internal substeps unless otherwise noted.

\subsection{Example: 2D-to-2D Square-to-Circle Kirigami Model}
\begin{figure}[t!]
    \centering
    \includegraphics[width=\linewidth]{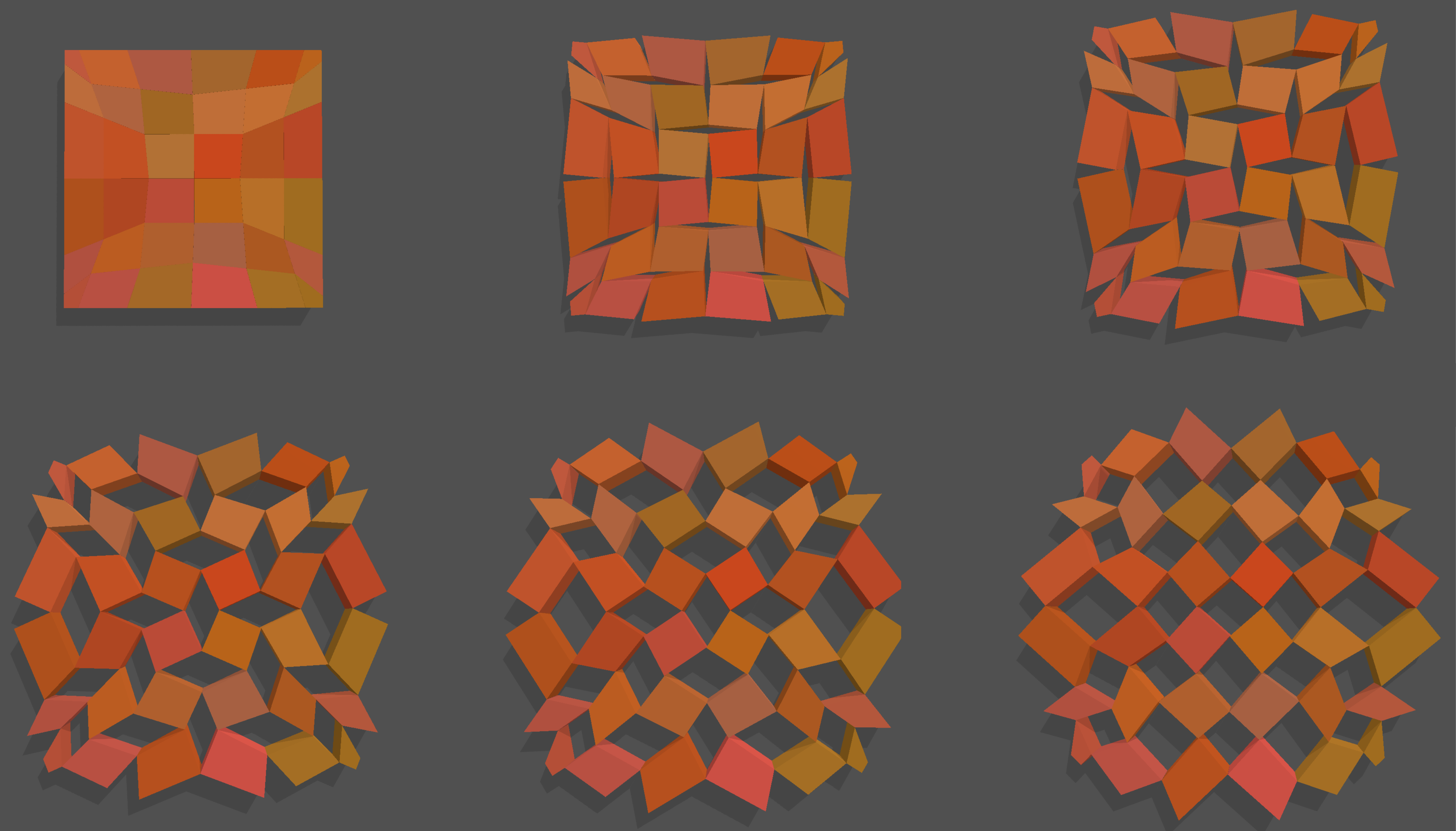}
    \caption{\textbf{Additional snapshots of the 2D-to-2D deployment of the square-to-circle kirigami model achieved by PyKirigami.}}
    \label{fig:si_sq2circ}
\end{figure}

The 2D-to-2D square-to-circle kirigami model example (Fig.~\ref{fig:si_sq2circ}; see also main text Fig.~3\textbf{a}) serves as an excellent test case for a 2D radial expansion because of the pattern symmetry. The primary challenge is to enforce planar motion and prevent the structure from buckling out-of-plane. Our strategy combines hinge kinematics with specific environmental and geometric settings, driven by an automated radial force model. 

\paragraph{Key Simulation Parameters}
A stable, planar, and automated deployment is achieved with the following configuration:
\begin{itemize}
    \item \textbf{Kinematic Model:} We use \textbf{hinge joints} to connect the tiles. This is the most important choice, as it constrains tile rotation to the Z-axis, fundamentally preventing out-of-plane tumbling.
    
    \item \textbf{Actuation Model:} The deployment is driven by the automated Center-of-Mass Expansion model (\texttt{--cm\_expansion}). No target file is needed.
    
    \item \textbf{Physical Environment:} \texttt{--gravity -10}, \texttt{--ground\_plane}

    \item \textbf{Geometry:} \texttt{--brick\_thickness 0.2}. This provides the tiles with rotational inertia to resist torsional instabilities.

\end{itemize}

\begin{figure}[t]
    \centering
    \includegraphics[width=1.0\linewidth]{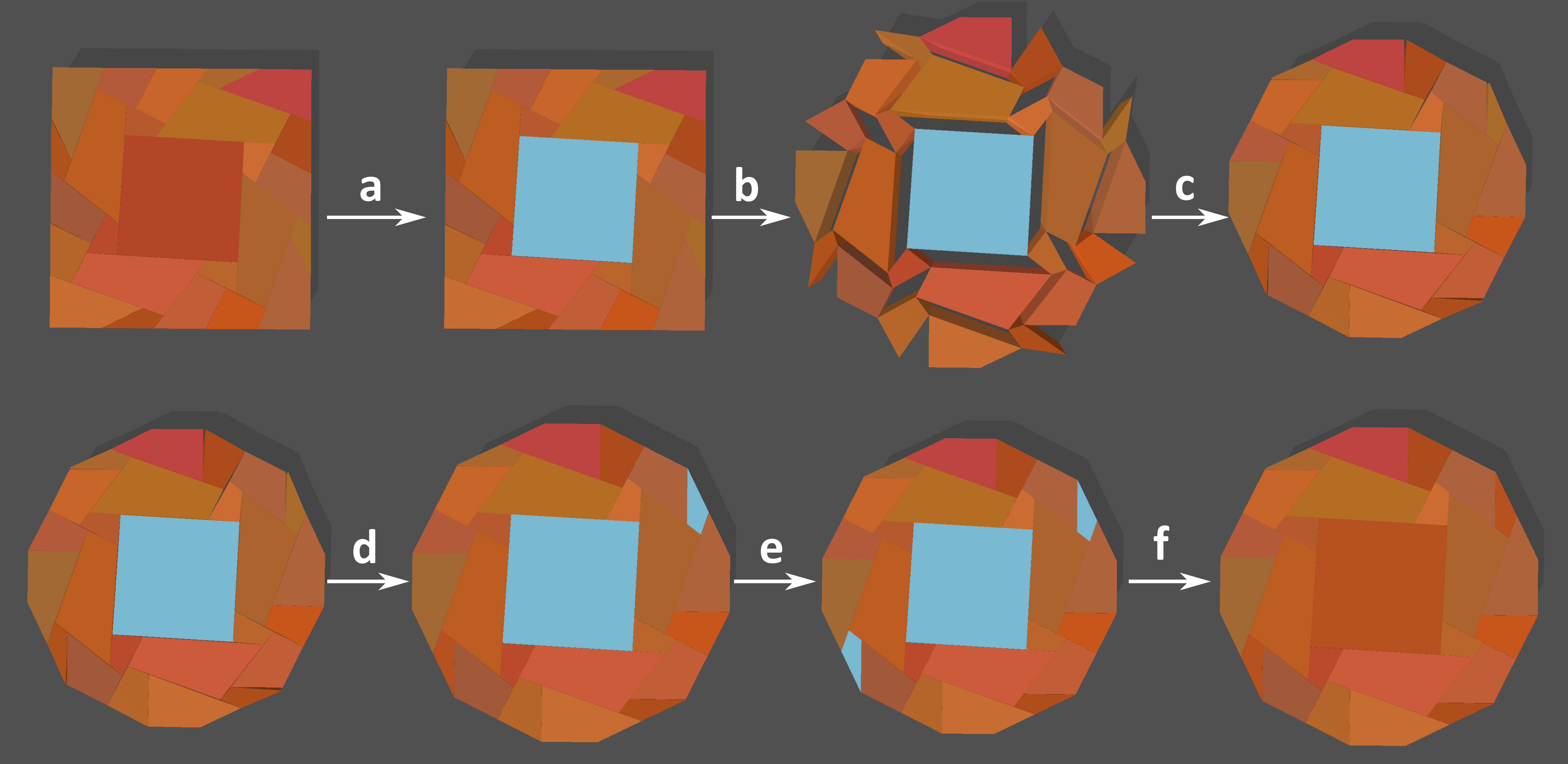}
    \caption{
            \textbf{Interactive workflow for the manual reconfiguration of the square-to-disk kirigami model.} 
            \textbf{a}, A central tile is pinned (\texttt{Right-click}) to create a stable reference frame.
            \textbf{b}, An outer tile is dragged (\texttt{Left-click + drag}) to begin the radial expansion.
            \textbf{c}, The structure is guided into an approximate ``disk'' shape. At this stage, small gaps often persist due to the implicit hinge forces.
            \textbf{d}, To close a gap, a tile is dragged into the correct position through \texttt{Left-click + drag} and the simulation is then paused (\texttt{Space bar}), and the tile is pinned (\texttt{Right-click}) while paused. This ``pause-and-pin'' technique is a powerful feature, allowing a constraint to be applied to a desired geometry before the physics solver can react.
            \textbf{e}, Step~\textbf{d} is repeated for other misaligned tiles.
            \textbf{f}, All pins are removed while paused, leaving the structure in a stable ``disk'' state.
    }
    \label{fig:si_square-to-disk}
\end{figure}

\subsection{Example: 2D-to-2D Square-to-Disk Kirigami Model (Interactive Workflow)} \label{subsec: si_square-to-disk}
Fig.~\ref{fig:si_square-to-disk} demonstrates the fully interactive workflow for the 2D-to-2D compact reconfigurable square-to-disk model shown in main text Fig.~3\textbf{b}. While the main text analyzes this structure's automated deployment, here we show how to achieve the same reconfiguration manually using the GUI controls detailed in Table~\ref{tab:controls}. This process highlights the use of the simulator for hands-on exploration and state capture.

With the simulation running in a no-force mode, a user can manually reconfigure the structure from the ``square'' to the ``disk'' state. This is achieved by applying the interactive controls from Table~\ref{tab:controls} in a strategic sequence, as illustrated in Fig.~\ref{fig:si_square-to-disk}.

\paragraph{Key Simulation Parameters:}
\begin{itemize}
     \item \textbf{No Scripted Actuation:} To enable manual control, we launch the simulation without any automated force models. This is done by omitting the \texttt{--cm\_expansion} flag and ensuring no \texttt{--target\_vertices\_file} is specified.
    \item \textbf{Physical Environment:} \texttt{--gravity -10} \texttt{--ground\_plane} 
    \item \textbf{Visualization}  \texttt{--brick\_thickness 0.2}.
\end{itemize}

\subsection{Example: 2D-to-3D Square-to-Spherical-Cap Kirigami Model}
The square-to-spherical-cap kirigami model example (Fig.~\ref{fig:si_partial_sphere}; see also main text Fig.~3\textbf{c}) demonstrates a 2D-to-3D transformation from a flat sheet into a curved surface. This complex, out-of-plane motion is impossible to guide with simple radial forces and necessitates the use of the target-based actuation model for precise control. Here, additional snapshots showing the progression from a planar to a spherical configuration are provided. Also, one can notice that out-of-plane motion provides extra degrees of freedom, allowing the sheet to lift/twist to resolve incompatibilities which can be seen in~\cite{choi2021compact}.

\begin{figure}[t!]
    \centering
    \includegraphics[width=\linewidth]{F13_SI_partial_sphere.pdf}
    \caption{\textbf{Additional snapshots of the 2D-to-3D deployment of the square-to-spherical-cap kirigami model achieved by PyKirigami, driven by the target-based force model.}}
    \label{fig:si_partial_sphere}
\end{figure}

\paragraph{Key Simulation Parameters:}
\begin{itemize}
    \item \textbf{Kinematic Model:} We use \textbf{spherical joints} to connect the bottom face of the tiles, allowing for the necessary three-dimensional rotational freedom.
    \item \textbf{Actuation Model:} The deployment is driven by the \textbf{Target-Based model}, with a target file defining the final spherical geometry.
    \item \textbf{Key Actuation Parameters:} \texttt{--spring\_stiffness 100}, \\\texttt{--force\_damping 50}. 
    \item \textbf{Physical Environment:} \texttt{--ground\_plane}. It can serve as a background to visualize the shadow.
    \item \textbf{Geometry:} \texttt{--brick\_thickness 0.02}. It is noteworthy that if the tiles are too thick, they will block the deployment.
\end{itemize}

\subsection{Example: 3D-to-3D Compact Reconfigurable Cylinder Kirigami Model}
\begin{figure}[t!]
    \centering
    \includegraphics[width=\linewidth]{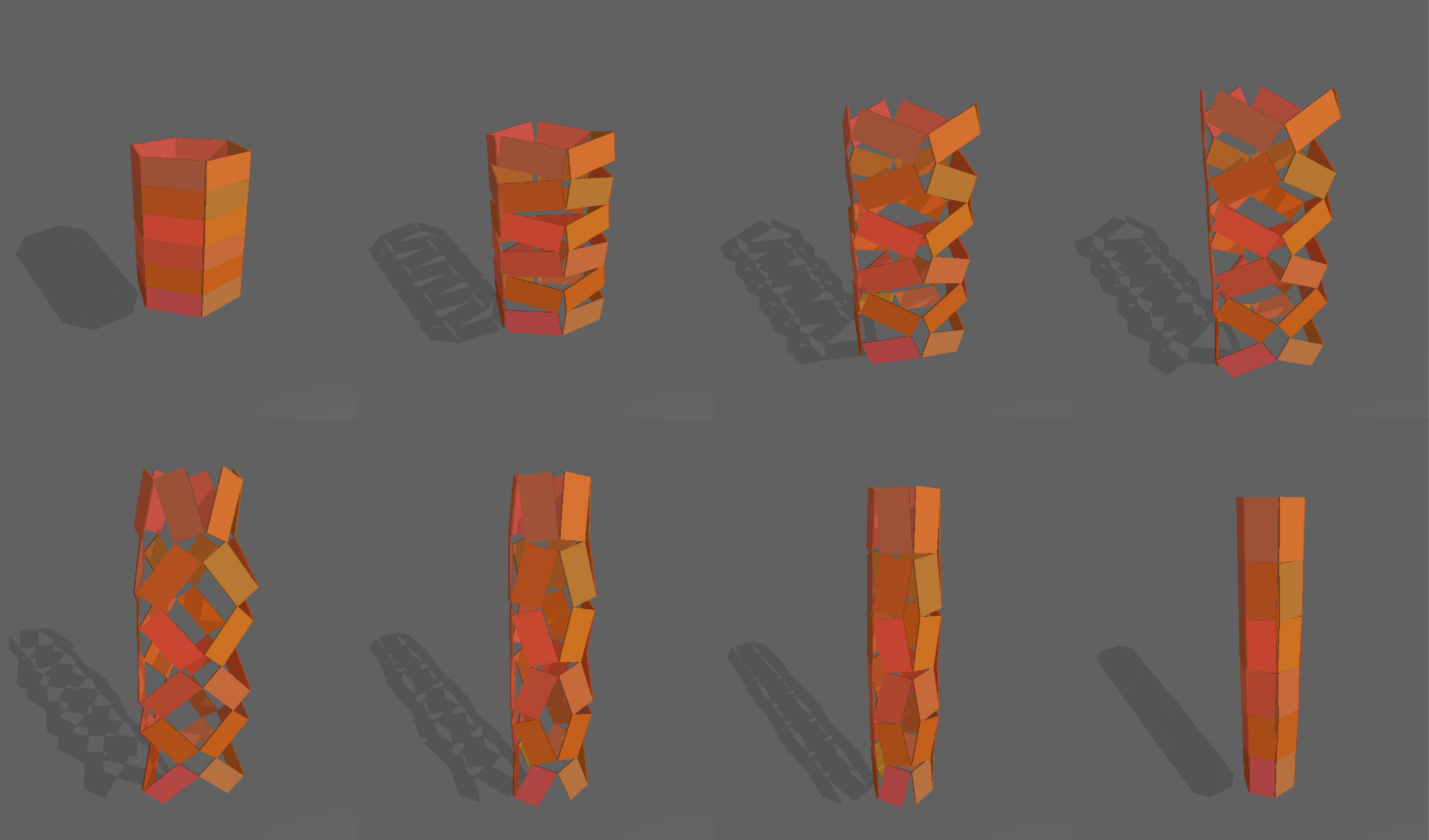}
    \caption{\textbf{Additional snapshots of the reconfiguration process of the 3D-to-3D cylinder kirigami structure achieved by PyKirigami.}}
    \label{fig:si_cylinder}
\end{figure}

The 3D-to-3D compact reconfigurable cylinder kirigami model example (Fig.~\ref{fig:si_cylinder}; see also main text Fig.~3\textbf{d}) involves the reconfiguration between two distinct 3D states, from a compact, folded cylinder to another compact, folded state. The target-based model is essential for guiding the structure through a collision-free volumetric transformation. 

\paragraph{Key Simulation Parameters:}
\begin{itemize}
    \item \textbf{Kinematic Model:} We use \textbf{spherical joints} to allow for full 3D motion between the tiles.
    \item \textbf{Actuation Model:} The reconfiguration is driven by the \textbf{Target-Based model}.
    \item \textbf{Key Actuation Parameters:} \texttt{--spring\_stiffness 100}, \\ \texttt{--force\_damping 80}.
    \item \textbf{Physical Environment:} \texttt{--gravity 0}
    \item \textbf{Geometry:}  \texttt{--brick\_thickness 0.02}. Note that thinner tiles yield a better deployment.
\end{itemize}

\subsection{Example: 2D-to-2D Stampfli Quasicrystal Kirigami Models}
Note that the above kirigami structures primarily consist of quadrilateral tiles, and the overall structures exhibit a highly regular topology. To further demonstrate the applicability of PyKirigami to other kirigami structures, we consider simulating the deployment of two quasicrystal kirigami structures from~\cite{liu2022quasicrystal} with highly different topologies.

\begin{figure}[t!]
    \centering
    \includegraphics[width=\linewidth]{F15_SI_stampfli24.pdf}
    \caption{\textbf{The simulated deployment of the 12-fold Stampfli quasicrystal kirigami structure from the Hamiltonian construction method in~\cite{liu2022quasicrystal}} achieved by PyKirigami with automated radial expansion, stabilized by environmental forces.}
    \label{fig:stampfli24}
\end{figure}

\begin{figure}[t!]
    \centering
    \includegraphics[width=0.96\linewidth]{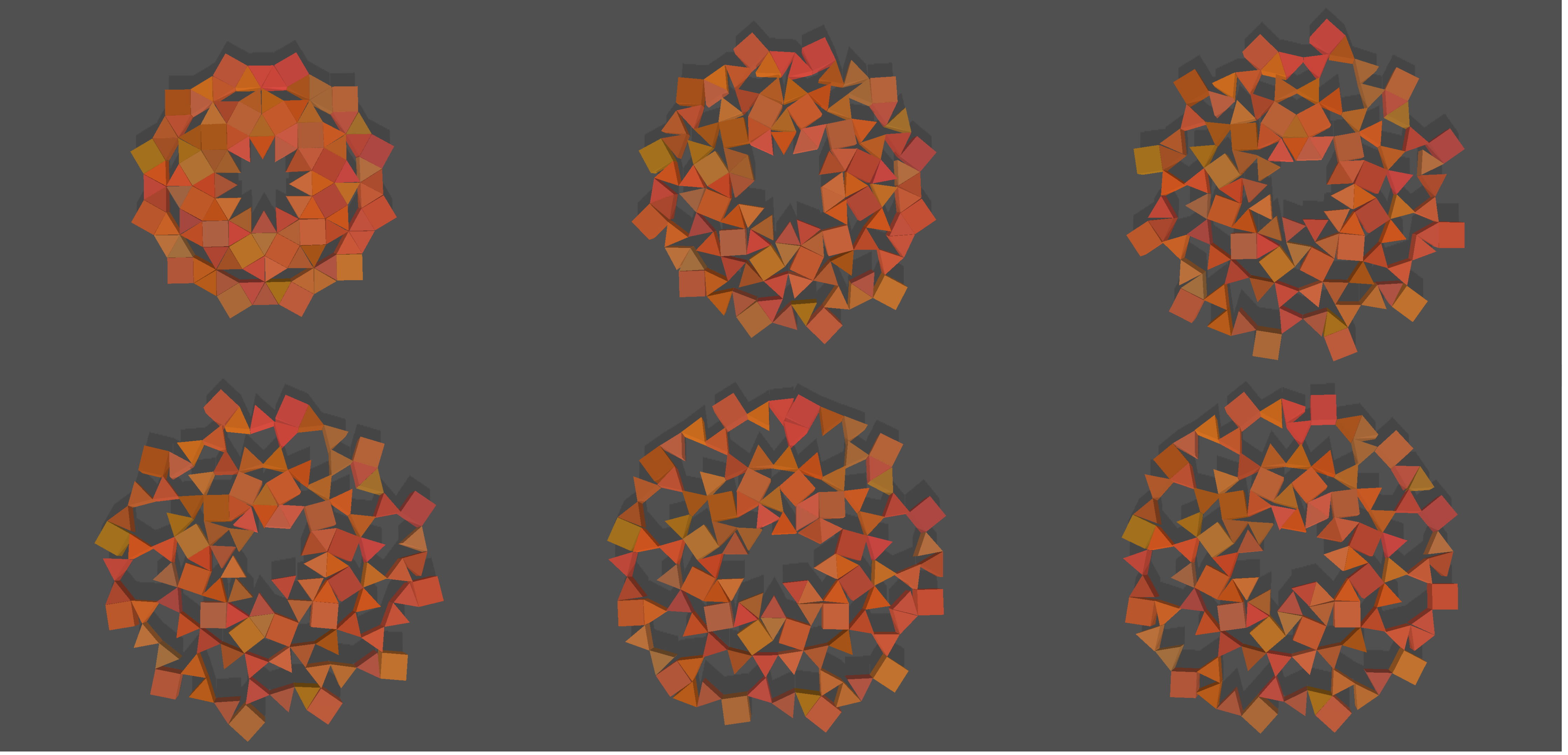}
    \caption{\textbf{The simulated deployment of the 12-fold Stampfli quasicrystal kirigami structure from the Tile Removal construction method in~\cite{liu2022quasicrystal} achieved by PyKirigami with automated radial expansion, stabilized by environmental forces.}}
    \label{fig:SI_stampfli132}
\end{figure}

More specifically, in Fig.~\ref{fig:stampfli24} we show the deployment of a 12-fold Stampfli quasicrystal kirigami structure created using the Hamiltonian construction method in~\cite{liu2022quasicrystal}. Here, note that the structure contains not only quadrilateral tiles but also triangular tiles, and the tile connection is also highly different from the ones shown above, with all tiles forming one big closed loop. It can be observed that PyKirigami is capable of producing a deployment simulation of this structure with automated radial expansion.

In Fig.~\ref{fig:SI_stampfli132}, we consider another 12-fold Stampfli quasicrystal kirigami structure created using the Tile Removal construction method in~\cite{liu2022quasicrystal}, which consists of both square and triangular tiles but possesses a highly different topology. Specifically, it contains a large hole at the center as well as multiple holes in the surrounding regions with different sizes and shapes. The successful deployment achieved by PyKirigami showcases the capability of PyKirigami in handling kirigami structures with different topologies.

\begin{figure}[t!]
    \centering
    \includegraphics[width=0.96\linewidth]{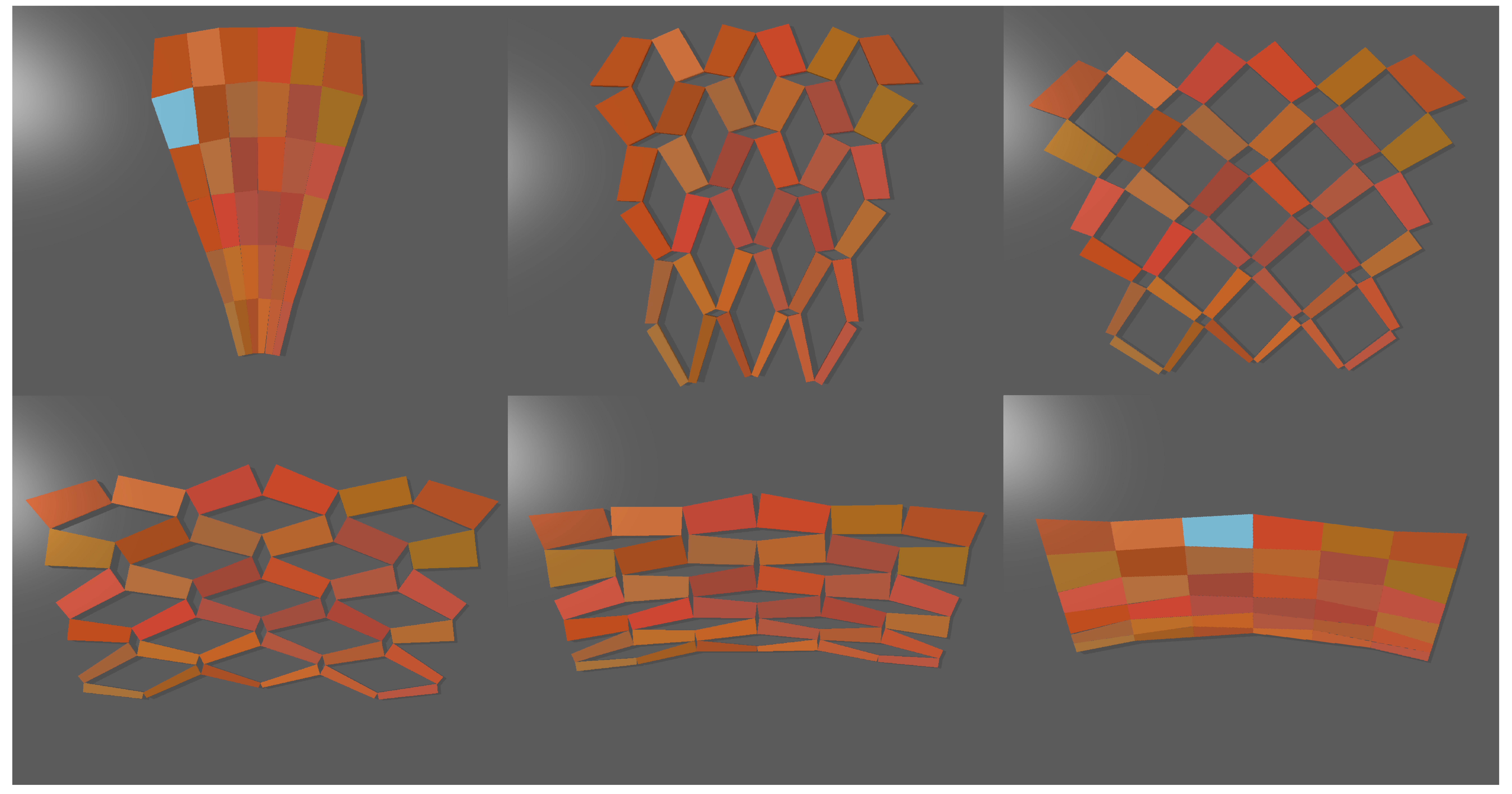}
    \caption{\textbf{The simulated deployment of the compact reconfigurable fan kirigami model produced by PyKirigami with interactive shape morphing.}}
    \label{fig:fan}
\end{figure}

\subsection{Example: 2D-to-2D Compact Reconfigurable Fan Kirigami Model}
To further demonstrate the versatility of PyKirigami, we present another example focusing on its interactive reconfiguration functionality. Fig.~\ref{fig:fan} shows the compact reconfigurable fan kirigami model, which has two different closed and compact contracted states. Users can apply the mentioned interactive operations to deploy it under the same physical environment as the square-to-disk model as shown in~\ref{subsec: si_square-to-disk}.

\subsection{Example: 3D-to-3D cube-to-sphere Kirigami Model}
\begin{figure}[t!]
    \centering
    \includegraphics[width=0.96\linewidth]{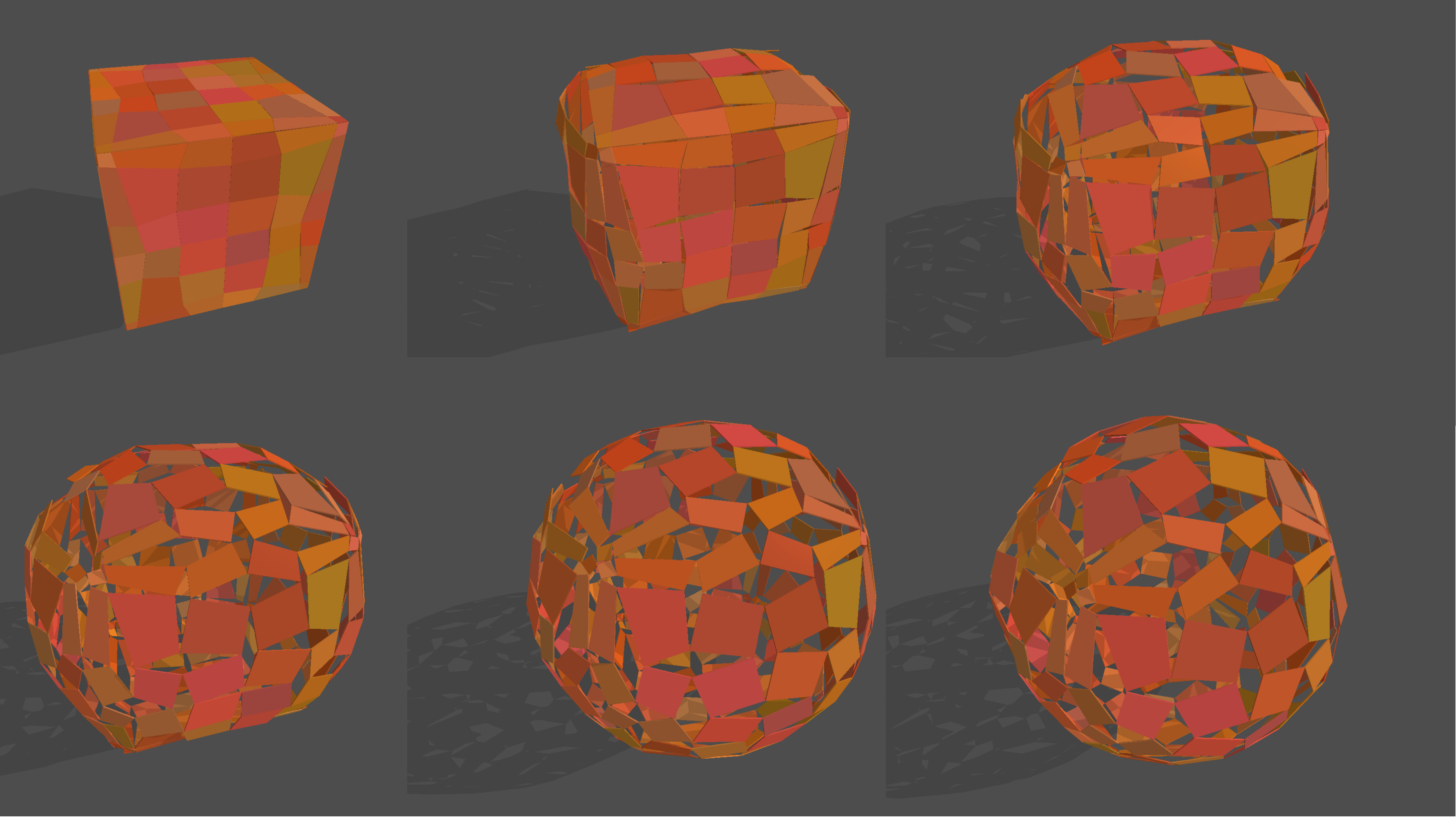}
    \caption{\textbf{The simulated deployment of the cube-to-sphere kirigami model produced by PyKirigami.} In this simulation, large, coordinated out-of-plane rotations and precise positioning are applied to control the deployment effect.}
    \label{fig:si_cube-to-sphere}
\end{figure}

Finally, we present an interesting 3D-to-3D transformation example using the cube-to-sphere kirigami model (Fig.~\ref{fig:si_cube-to-sphere}; see also main text Fig.~1\textbf{b}). For this model, the six faces of the cube are deployed into a spherical approximation throughout the deployment process, which requires large, coordinated out-of-plane rotations and precise positioning to ensure the edges meet correctly without self-intersection. Such a transformation is infeasible with simple or manual forces and serves as a powerful demonstration of the target-based actuation model's capabilities.

\subsection{Handling Non-Rigidly Deployable Structures} \label{appendix:complex}

While PyKirigami primarily considers kirigami tiles as 3D rigid bodies and focuses on rigid deployments, it can also be utilized for handling kirigami structures that are not rigidly deployable.
 
\begin{figure}[t]
    \centering
    \includegraphics[width=1.0\linewidth]{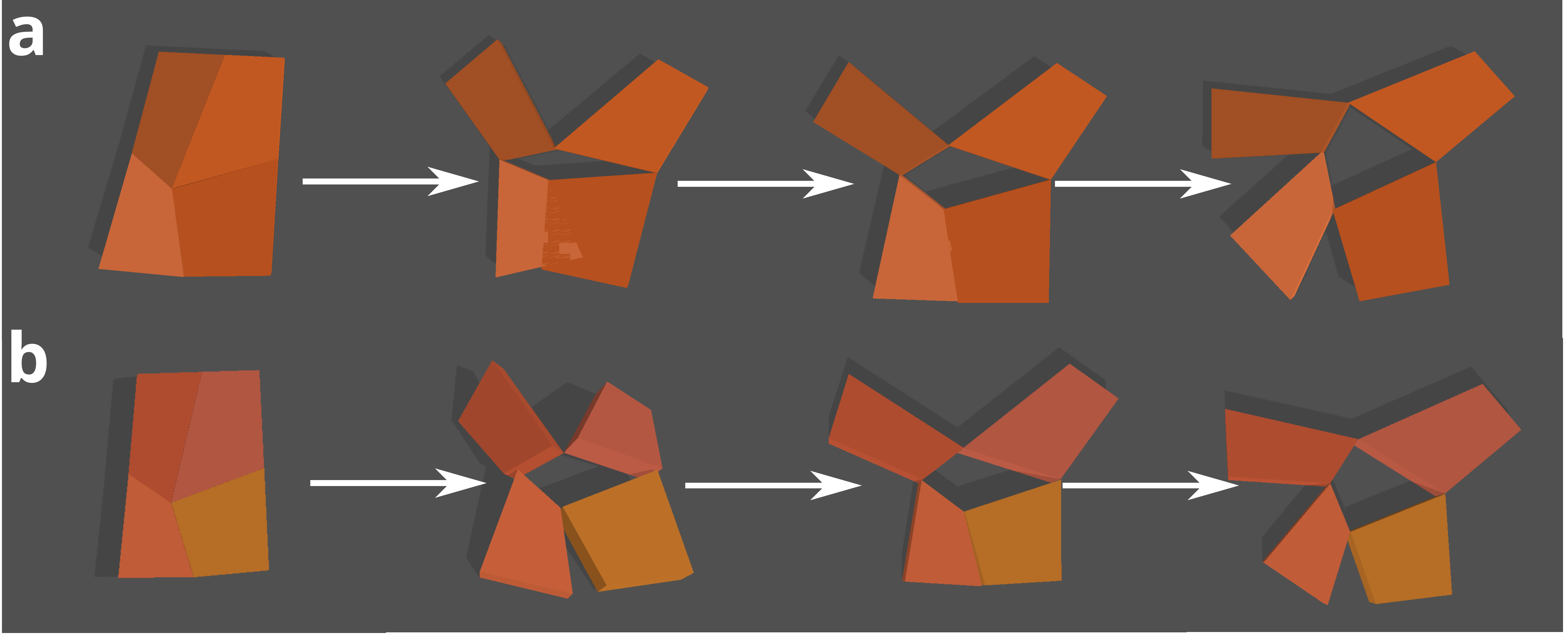}
    \caption{\textbf{Deployment of a non-rigidly-deployable kirigami structure.} (a) Planar rigid deployment is geometrically impossible due to tile overlap. (b) By utilizing the 3D deployment ability in PyKirigami, the incompatibility can be resolved: the unit deploys out of plane (second frame), creating nonplanar negative space, then flattens back with the introduced gaps (third frame).}
    \label{fig:demo_nonrigid}
\end{figure}

Specifically, to handle the deployment of non-rigidly deployable planar kirigami structures, one strategy is to utilize the additional degree of freedom provided in the spatial dimension. For instance, Fig.~\ref{fig:demo_nonrigid} illustrates a simple 2D non-rigidly deployable kirigami structure consisting of four quadrilaterals. As the slit at the interior does not form a straight line, this pattern is not rigidly deployable in 2D theoretically~\cite{choi2021compact}. Panel (a) shows that attempting to deploy this pattern in the plane leads to tile overlap, suggesting that a strictly planar deployment is geometrically impossible. Nevertheless, since PyKirigami handles kirigami structures in the 3D space, we can easily handle such cases by allowing the structure to move out of plane as shown in Panel (b). In the second frame, the unit deploys in 3D space, where the additional degree of freedom resolves the geometric conflict. Crucially, the negative space (the gap between tiles) at this intermediate state is nonplanar, as the four tile centers do not lie in a single plane. This ``nonplanar negative space'' is a characteristic feature of 3D kirigami deployment and distinguishes it from purely 2D mechanisms. At the third frame, the structure is flattened back to the plane with negative spaces introduced, achieving a fully deployed planar configuration that was impossible in panel (a).

More generally, to model non-rigid deployments of kirigami structures in either 2D or 3D, the user may consider a subdivision of a tile into smaller units and connect them appropriately, analogous to the use of a triangulated origami mesh in prior non-rigid origami folding simulators~\cite{tachi2010freeform,liu2016merlin,liu2018highly,ghassaei2018fast}. For instance, the bending of a kirigami tile can be modelled by replacing the tile with two or multiple smaller tiles connected by hinge joints along the tile edges, which provides additional flexibility in simulating 3D kirigami deployment. Similarly, the stretching and deformation of a tile can be modelled by replacing it with multiple smaller tiles connected by joints, so that the smaller tiles together with the negative spaces formed during the deployment can serve as a representation of the tile stretching and deformation.

\subsection{Handling More Complex Deployments} \label{appendix:complex_deployment}

\begin{figure}[t!]
    \centering
    \includegraphics[width=\linewidth]{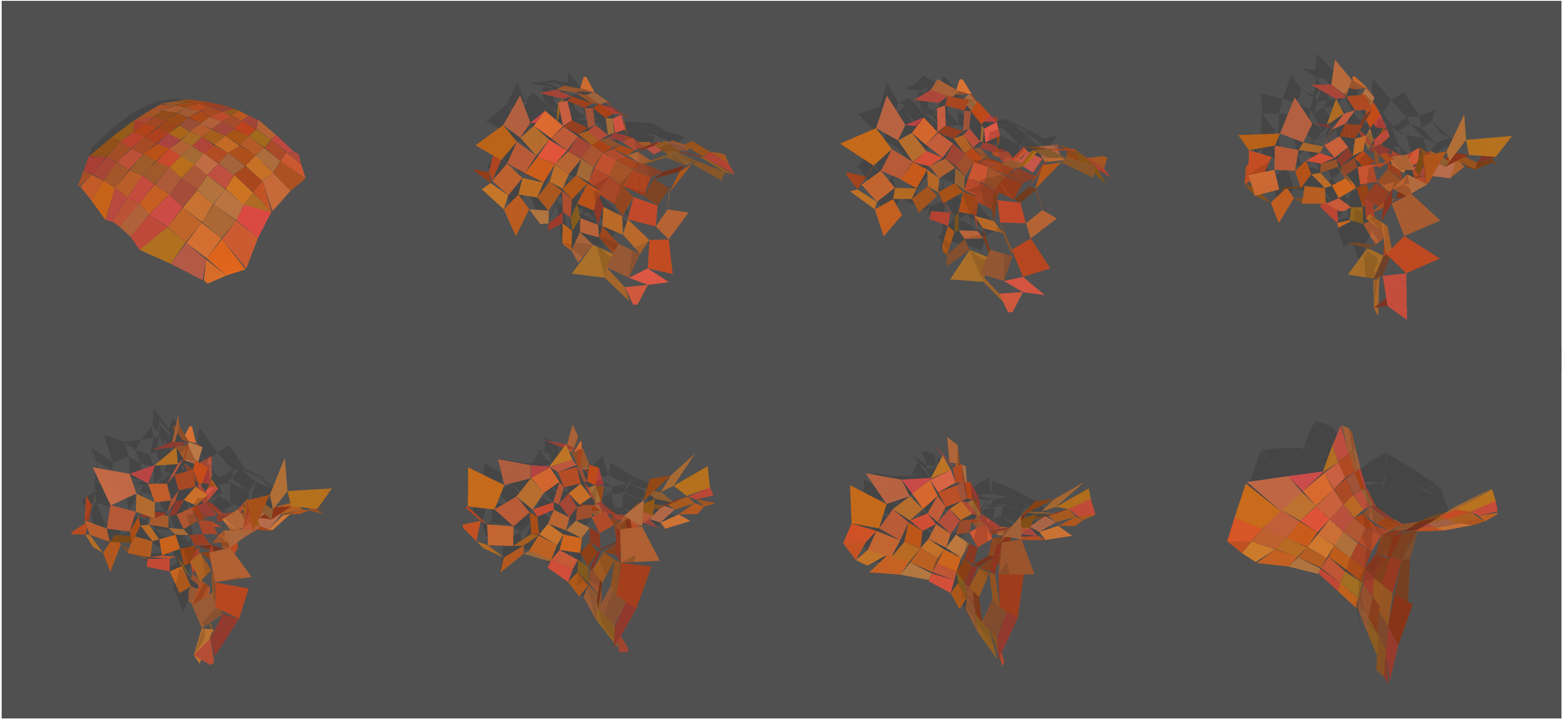}
    \caption{\textbf{The simulated deployment of the sphere cap to saddle produced by PyKirigami.} Automatic connection-type detection assigns 135 bottom and 45 top joints based on the saddle target geometry. A very small brick thickness (0.001) renders initial constraint violations negligible.}
    \label{fig:si_sphere-to-saddle}
\end{figure}

As a challenging test case, we simulate the deployment of a kirigami structure with 100 tiles and 180 joints from a compact spherical-cap shape to a compact saddle shape (Fig.~\ref{fig:si_sphere-to-saddle}). It is noteworthy that the initial and target configurations exhibit significantly different surface curvatures, making the deployment process challenging. Nevertheless, PyKirigami is capable of producing a real-time deployment that runs interactively at 240 HZ with adaptive stiffness.

Note that the interaction between thickness and connection type introduces a further consideration for 3D kirigami deployment to curved targets. Ideal kirigami tiles are planar polygons without thickness, and in that case the choice of ``mountain fold'' (bottom connection) or ``valley fold'' (top connection) does not matter. However, for approximating a curved target using kirigami structures with tile thickness, the connection type plays an important role. Using the automatic connection-type detection (\texttt{--auto\_detect\_connections}) with the saddle as the target geometry, PyKirigami assigns 135 bottom-face joints and 45 top-face joints based on the surface curvature of the target geometry to facilitate the deployment. Also, since we form 3D tile bricks by extruding the planar tile and choosing the original planar tile as the bottom face, it is natural to take the thickness into account and analyze the numerical error when the connection type is 2, i.e., a joint connecting the top faces between two tiles. In this case, the distance between the corresponding bottom vertices is given by
$$
\epsilon(h, \theta) = \|v_{i,p} - v_{j,q}
\| = 2h\cos(\theta/2),
$$
where $v^{t}_{i,p}=v^{t}_{j,q}$ are enforced by constraint tuple $(i,p,j,q,2)$ and $\theta$ is the dihedral angle of valley fold. This error is already incorporated in the maximal deviation from the target, and one can reduce this part by selecting a small thickness. To handle the deployment of this kirigami structure, we use a very small brick thickness ($h=0.001$).

\begin{figure}[t]
    \centering
    \includegraphics[width=1.0\linewidth]{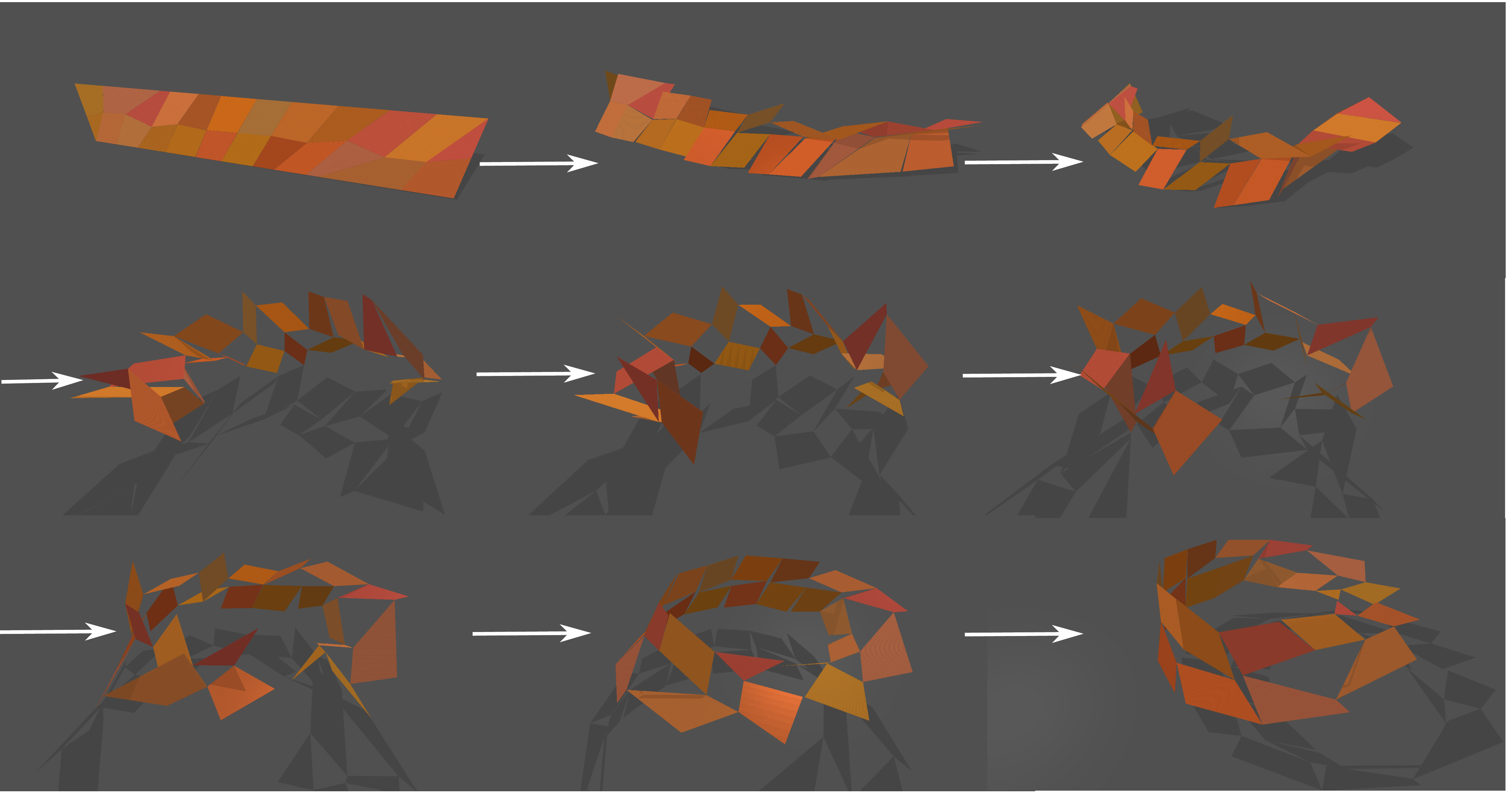}
    \caption{\textbf{3D deployment sequence of a 24-tile kirigami structure from a rectangular band to a M\"obius strip.} 
    The 3$\times$3 frame sequence illustrates the full shape-morphing process from the planar rectangular configuration to a topologically non-trivial single-sided M\"obius configuration.}

    \label{fig:demo_mobius}
\end{figure}

Next, we further demonstrate the effectiveness of PyKirigami in handling deployments involving structural twists and orientation changes. Fig~\ref{fig:demo_mobius} shows the deployment of a kirigami structure from a rectangular band to a M\"{o}bius strip. Specifically, note that the underlying shapes at the two states have different topologies, and another challenging aspect is the change of the outward surface normal direction of the tiles throughout the deployment. As shown in the simulation results, PyKirigami successfully produces a collision-free deployment path to the single-sided target M\"{o}bius strip geometry.

Altogether, these examples demonstrate the capability of PyKirigami in achieving desired complex shape-morphing effects.

\section{Analytic Kinematic Models for Quantitative Validation} \label{appendix:analytic}

\begin{figure}[t]
    \centering
    \includegraphics[width=0.9\linewidth]{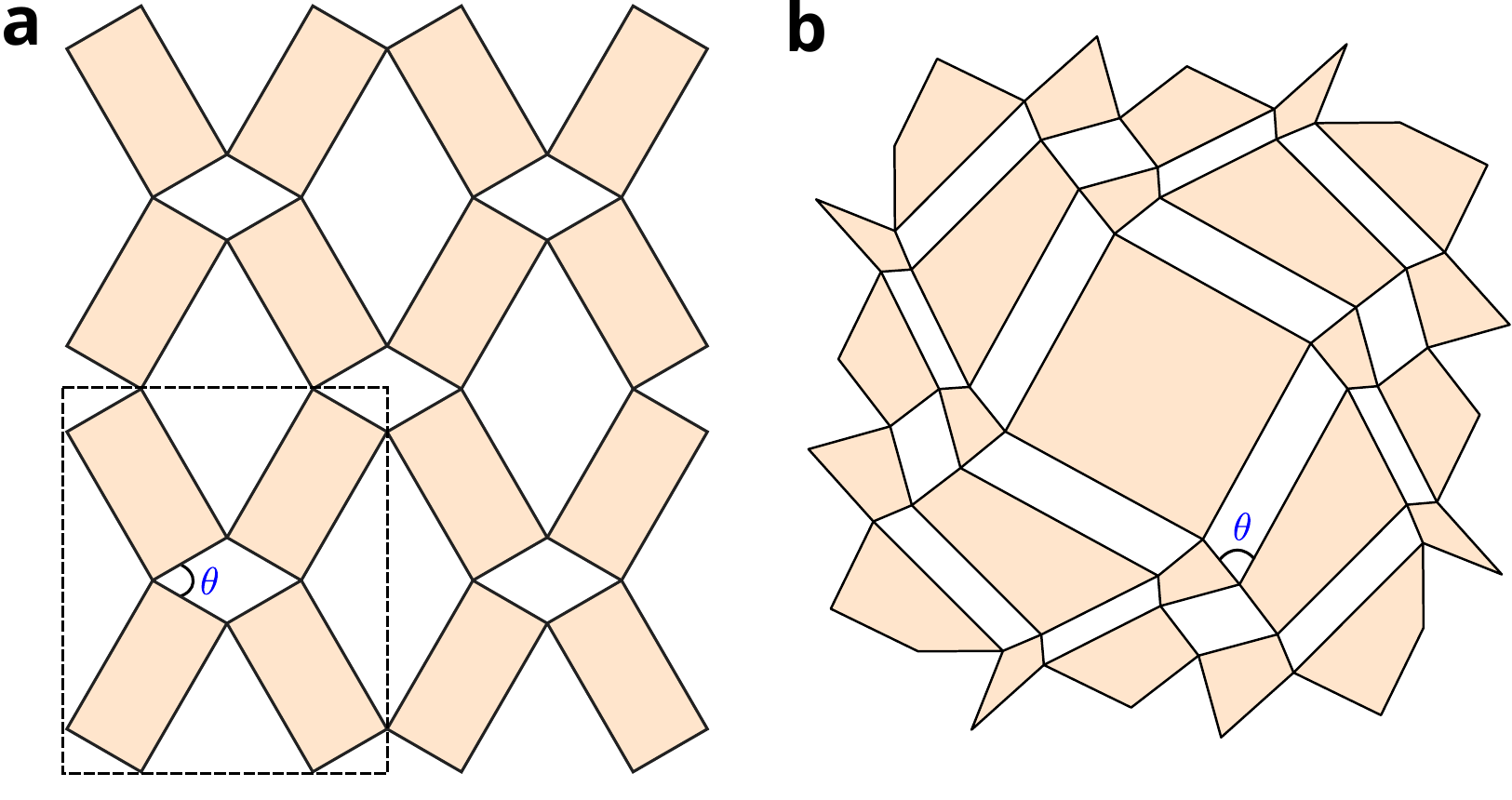}
    \caption{\textbf{The definition of the deployment angle in the benchmark cases.}\textbf{a}. The rectangular tessellation kirigami structure. The dashed line shows a unit cell of the tessellation, in which the deployment angle $\theta$ can be defined. \textbf{b}, The rigidly-deployable square-to-disk kirigami structure from~\cite{dudte2023additive}, in which the deployment can be characterized by a single deployment angle $\theta$.}
    \label{fig:SI_analytic_demo}
\end{figure}

To validate the numerical accuracy of PyKirigami, we derive the exact kinematic equations for the two benchmark kirigami structures used in main text. Both systems function as single-degree-of-freedom (1-DOF) mechanisms, where the global configuration is uniquely determined by the deployment angle $\theta$.

\subsection{Rectangular Tessellation}
The rectangular tessellation is frequently used as a base case in kirigami design, and its analytical vertex position along with deployment angle $\theta$ can be easily derived. As shown in Fig.~\ref{fig:SI_analytic_demo}\textbf{a}, the unit cell consisting of four rectangles has the size $L\times W$ where
$$
\begin{pmatrix}
    L\\
    W
\end{pmatrix} = 2 \begin{pmatrix}
    \sin(\eta) & \cos(\eta)\\
    \cos(\eta) & \sin(\eta)
\end{pmatrix}\begin{pmatrix}
    l\\
    w
\end{pmatrix}, \quad \eta = \theta/2,
$$
assuming that the small rectangles have size $l\times w$.

\subsection{Square-to-Disk Kirigami Structure}
The compact reconfigurable square-to-disk kirigami structure exhibits a more complex, non-linear kinematic path. The radial position $R$ of boundary vertices is coupled to the deployment angle $\theta$ (Fig.~\ref{fig:SI_analytic_demo}\textbf{b}). The position of all inner vertices $\mathbf{v}_{k}$ in the assembly can be derived via a matrix multiplication
$$
\mathbf{v}_{k} = M V,
$$
where the matrix $M$ is related to the deployment angle and $V$ are the seed vertices. More details can be found in~\cite{dudte2023additive}.

\section{Software Environment and Reproducibility} \label{appendix:reproducibility}
\subsection{Software Environment}
PyKirigami has been developed and tested on the platforms 
and software versions listed in Table~\ref{tab:environment}. The experiments to analyze sensitivity and compare PyKirigami and Pymunk are implemented on Macbook Air (M1). 

\begin{table}[h!]
\centering
\caption{\textbf{Tested software environment.}}
\label{tab:environment}
\begin{tabularx}{\linewidth}{@{} l l l @{}}
\toprule
\textbf{Component} & \textbf{Tested version} & \textbf{Notes} \\
\midrule
Python       & 3.13  & \texttt{python -V} to check \\
NumPy        & 2.4.1         & \texttt{conda list numpy} \\
PyBullet     & 3.25        & \texttt{conda list pybullet} \\
\midrule
\multirow{3}{*}{Operating system}
             & Ubuntu 24.04 LTS & fully tested \\
             & macOS 26 (Tahoe) & fully tested \\
             & Windows 11   & GUI tested; headless supported \\
\bottomrule
\end{tabularx}
\end{table}

\subsection{Reproduction Commands for All Examples}
Table~\ref{tab:commands} lists the exact commands used to produce every result in the paper.  All required input files are included in the \texttt{data/} directory of the repository. The \texttt{{-}{-}headless} flag may be appended to any command to run without a GUI for automated batch execution.

\begin{table}[h!]
\small
\centering
\caption{\textbf{Reproduction commands for all examples.}
All commands are run from the repository root.  Only non-default flags are shown; defaults follow Table~\ref{tab:si_cli_params}.  Append \texttt{--headless} to any command for batch execution without a GUI.}
\label{tab:commands}
\begin{tabularx}{\linewidth}{@{} X @{}}
\toprule

\textit{\textbf{2D deployments}} \\[3pt]

\rowcolor{Gray}
Square-to-circle \hfill \textit{Fig.~\ref{fig:results_combined}\textbf{a}, Fig.~\ref{fig:si_sq2circ}} \\
{\small\texttt{python run\_sim.py --model square2circle\_w3\_h3 --brick\_thickness 0.2 --spring\_stiffness 300 --ground\_plane --gravity -10 --cm\_expansion}} \\[4pt]

\rowcolor{Gray}
Square-to-disk \hfill \textit{Fig.~\ref{fig:results_combined}\textbf{b}, Fig.~\ref{fig:benchmark_sq2disk}, Fig.~\ref{fig:si_square-to-disk}} \\
{\small\texttt{python run\_sim.py --model square2disk --brick\_thickness 0.2 --spring\_stiffness 25 --ground\_plane --gravity -10 }} \\[4pt]

\rowcolor{Gray}
Stampfli-24 quasicrystal \hfill \textit{Fig.~\ref{fig:stampfli24}} \\
{\small\texttt{python run\_sim.py --model stampfli24 --brick\_thickness 0.2 --ground\_plane --gravity -10 --cm\_expansion}} \\[4pt]

\rowcolor{Gray}
Stampfli-132 quasicrystal \hfill \textit{Fig.~\ref{fig:SI_stampfli132}} \\
{\small\texttt{python run\_sim.py --model stampfli132 --brick\_thickness 0.2 --ground\_plane --gravity -10 --cm\_expansion}} \\[4pt]

\rowcolor{Gray}
Fan model \hfill 
\textit{Fig.~\ref{fig:fan}} \\
{\small\texttt{python run\_sim.py --model fan --brick\_thickness 0.2 --ground\_plane --gravity -10}} \\[4pt]

\rowcolor{Gray}
Nonrigid Demo \hfill 
\textit{Fig.~\ref{fig:demo_nonrigid}\textbf{a}} \\
{\small\texttt{python run\_sim.py --model demo\_nonrigid --brick\_thickness 0.2 --ground\_plane --gravity -10 --filter\_collision}} \\[4pt]

\midrule

\textit{\textbf{3D deployments}} \\[3pt]

\rowcolor{Gray}
Square-to-spherical-cap \hfill \textit{Fig.~\ref{fig:results_combined}\textbf{c}, Fig.~\ref{fig:si_partial_sphere}} \\
{\small\texttt{python run\_sim.py --model partialSphere}} \\[4pt]

\rowcolor{Gray}
Cylinder \hfill \textit{Fig.~\ref{fig:results_combined}\textbf{d}, Fig.~\ref{fig:si_cylinder}} \\
{\small\texttt{python run\_sim.py --model cylinder --force\_damping 80}} \\[4pt]

\rowcolor{Gray}
Cube-to-sphere \hfill \textit{Fig.~\ref{fig:si_cube-to-sphere}} \\
{\small\texttt{python run\_sim.py --model cube2sphere\_w3\_h3}} \\[4pt]

\rowcolor{Gray}
Nonrigid Demo \hfill \textit{Fig.~\ref{fig:demo_nonrigid}\textbf{b}}\\
{\small\texttt{python run\_sim.py --model demo\_nonrigid --brick\_thickness 0.2 --ground\_plane --gravity -10}} \\[4pt]

\rowcolor{Gray}
Sphere-to-saddle \hfill \textit{Fig.~\ref{fig:si_sphere-to-saddle}} \\
{\small\texttt{python run\_sim.py --model sphere-to-saddle --brick\_thickness 0.001 --auto\_detect\_connections}} \\[4pt]

\rowcolor{Gray}
M\"{o}bius strip \hfill \textit{Fig.~\ref{fig:demo_mobius}} \\
{\small\texttt{python run\_sim.py --model mobius --brick\_thickness 0.005 --spring\_stiffness 25 --torque\_stiffness 25}} \\[4pt]

\bottomrule
\end{tabularx}
\end{table}

\section{Video Captions}
\textbf{Supplementary Video 1}: The 2D-to-2D deployment of the square-to-disk kirigami model achieved by PyKirigami.

\textbf{Supplementary Video 2}: The 2D-to-3D deployment of the square-to-spherical-cap kirigami model achieved by PyKirigami.

\textbf{Supplementary Video 3}: The 3D-to-3D deployment and reconfiguration of the cylinder kirigami model achieved by PyKirigami.

\textbf{Supplementary Video 4}: The 3D-to-3D deployment of the cube-to-sphere kirigami model achieved by PyKirigami.

\end{document}